\documentclass[a4paper,11pt]{article}
\pdfoutput=1
\usepackage{jheppub}
\usepackage{slashed,subcaption,amssymb,amsmath,bm}
\usepackage[utf8]{inputenc}
\usepackage{amsfonts}
\usepackage{mathtools}
\usepackage[dvipsnames]{xcolor}
\usepackage{forloop}

\usepackage{ifpdf}
\graphicspath{{./img/}}
\usepackage{hyperref}
\newcommand{\gev}{\text{GeV}}
\newcommand{\tev}{\text{TeV}}

\newcommand{\me}{\mathrm{e}}

\DeclarePairedDelimiter\abs{\lvert}{\rvert}%
\makeindex

\title{Exploring Fine-tuning of the Next-to-Minimal Composite Higgs Model}

\author[a]{Daniel Murnane,}
\author[a]{Martin White,}
\author[a]{and Anthony G. Williams}

\affiliation[a]{ARC Centre of Excellence for Particle Physics at the Terascale, Department of Physics, University of Adelaide, South Australia 5005, Australia}

\emailAdd{daniel.murnane@adelaide.edu.au, ORCID: 0000-0003-4046-4822}
\emailAdd{martin.white@adelaide.edu.au}
\emailAdd{anthony.williams@adelaide.edu.au, ORCID: 0000-0002-1472-1592}

\abstract{We perform a detailed study of the fine-tuning of the two-site, 4D, Next-to-Minimal Composite Higgs Model (NMCHM), based on the global symmetry breaking pattern $SO(6)\rightarrow SO(5)$. Using our previously-defined fine-tuning measure that correctly combines the effect of multiple sources of fine-tuning, we quantify the fine-tuning that is expected to result from future collider measurements of the Standard Model-like Higgs branching ratios, in addition to null searches for the new resonances in the model. We also perform a detailed comparison with the Minimal Composite Higgs Model, finding that there is in general little difference between the fine-tuning expected in the two scenarios, even after measurements at a high-luminosity, 1 TeV linear collider. Finally, we briefly consider the relationship between fine-tuning and the ability of the extra scalar in the NMCHM model to act as a dark matter candidate, finding that the realisation of a $Z_2$ symmetry that stabilises the scalar is amongst the most natural solutions in the parameter space, regardless of future collider measurements.}

\keywords{Technicolor and Composite Models, Beyond Standard Model, Effective Field Theories, Global Symmetries}
\arxivnumber{1810.08355}

\date{}

\begin{document}

\maketitle

\section{Introduction}

Models that extend the Standard Model (SM) to include a composite sector are a popular way of naturalising the hierarchy between the observed Higgs mass and the mass one would expect to be generated from quantum loop corrections. The underlying composite behaviour is expected to be described with non-perturbative physics, but below a certain energy scale $\Lambda_\textnormal{composite}$ the physics confines, and the system can be described by an effective field theory (EFT) in which the Higgs emerges as a pseudo-Nambu-Goldstone boson of a spontaneously-broken global symmetry of the composite sector. The simplest model that is consistent with custodial symmetry, whilst leading to exactly four Nambu-Goldstone bosons, is based on the symmetry breaking pattern $SO(5)\times U(1)_X \rightarrow SO(4) \times U(1)_X$~\cite{contino2006,contino2007b}, which leads to the Minimal Composite Higgs Model (MCHM). The EW group $SU(2)\times U(1) \in SO(4) \equiv SU(2)\times SU(2)$ is gauged, giving rise to a naturally light (relative to the symmetry breaking scale) pseudo-Nambu-Goldstone-Boson (pNGB) Higgs.

The precise mechanics of this symmetry breaking have been explored in various contexts: studies with no assumptions of the higher scale \cite{kaplan1984,kaplan1984b,kaplan1985,kaplan1991,Barnard:2015ryq}, simple assumptions of fundamental heavy fermions to give rise to the composite sector \cite{sannino2004,sannino2014}, and the composite sector as arising from extra-dimensional effects \cite{contino2003,agashe2005,contino2006,contino2007b,giudice2007,contino2007,Dillon:2018wye}. We will work in the multi-site effective field formalism known as the 4D Composite Higgs Model (4DCHM), a thorough review of which can be found in \cite{Panico:2015jxa}. In this multi-site approach, the non-Higgs SM fields are treated as elementary, while the Higgs and heavy composite fields are grouped into discrete sectors, effectively obeying non-linear sigma models. Coupling between sectors is achieved with Yukawa-type interactions, and linear couplings of elementary and composite fermions, leading to partial compositeness of the physical fermions.

While this extension to the SM does provide a natural cut-off to the mass-generating self-corrections of the Higgs, the various incarnations still require some degree of fine-tuning to remain compatible with the observed SM-like Higgs boson mass and signal strengths, plus the lack of observed new particle content at the LHC. One can attempt to reduce the tuning by expanding the composite sector, for example by coupling the SM to more than one site of composite quark partners \cite{panico2012,de2012}; considering the leptons as partially composite, leading to naturalness by accidental cancellations \cite{carmona2015,Barnard:2017kbb}; or expanding the set of symmetries obeyed by the composite sector \cite{gripaios2009beyond,Redi:2012ha,Banerjee:2017qod}. In previous work, we have performed comprehensive scans of a variety of MCHMs (distinguished by different fermion embeddings, and different choices for which fermions are partially composite)~\cite{Barnard:2015ryq,Barnard:2017kbb}. In each case, the regions of the parameter space consistent with the Higgs VEV, top quark mass and the Higgs mass were identified and used to obtain current and projected constraints on fine-tuning as a function of existing and hypothetical limits on the top partner masses, charged vector boson resonance masses, Higgs coupling deviations and the compositeness scale. We also presented measures of fine-tuning that accurately count the variety of higher order tunings that exist in composite Higgs theories, which can result from, for example, tuning the parameters to obtain a Higgs VEV below the compositeness scale, and then separately tuning them to ensure that leading and sub-leading contributions to the Higgs potential are sufficiently matched to break electroweak symmetry.

In this paper, we extend our earlier results to perform a detailed comparison of the fine-tuning of the MCHM and its minimal extension, the Next-to-Minimal Composite Higgs Model (NMCHM) based on the symmetry breaking pattern $SO(6) \rightarrow SO(5)$. This introduces an extra $SO(4)$-singlet scalar along with the four components of the usual Higgs doublet. We also include composite fermions in order to render the radiatively generated Higgs potential finite, and it can be shown using Weinberg sum rules that the minimal number of composite fermions required is two \cite{Banerjee:2017qod}. Previous works have thoroughly constructed and explored the naturalness of the two-site NMCHM, both effectively \cite{niehoff2017electroweak} and with UV completions in the fundamental partial compositeness paradigm \cite{BuarqueFranzosi:2018eaj}. However, its phenomenology has only been tested against naive measures of tuning, using non-convergent scanning techniques. In this paper, we focus on those qualities of the NMCHM that differentiate its higher-order tuning from the MCHM, and explore both models using a novel scanning technique called ``differential evolution" (DE), that allows us to obtain convergent results where other techniques have previously failed. 

This paper is structured as follows. In Section~\ref{model}, we review the NMCHM, before giving the details of our scanning procedure in Section~\ref{scanning}. We present our fine-tuning measure and results in Section~\ref{results}, including a comparison of the MCHM and NMCHM results, and a discussion of the potential of the scalar singlet to act as a dark matter candidate. Finally, we present our conclusions in Section~\ref{sec:conclusions}.

\section{The Next-to-Minimal Composite Higgs}
\label{model}



\subsection{Group structure}

For details on the formalism of the Minimal Composite Higgs Model (MCHM), we refer the reader to \cite{contino2003,agashe2005,contino2006,contino2007b,giudice2007,contino2007,Panico:2015jxa}. Here, we shall instead focus on what differentiates the Next-to-Minimal Composite Higgs Model (NMCHM) from the MCHM.
In particular, we use the two-site construction first described in \cite{de2012}, with its composite fermions and scalar resonances to render the pNGB Higgs potential finite. 



The five pNGBs from the spontaneous breaking of the global $SO(6)\rightarrow SO(5)$ symmetry are parameterised as:

\begin{align}
\Phi = e^{\frac{\sqrt{2}}{f}i\pi^{\hat{a}}(x)T^{\hat{a}}}\Phi_0 = \frac{1}{\varphi}\sin\frac{\varphi}{f}\left(h_1, h_2, h_3, h_4, s, \varphi \cot \frac{\varphi}{f}\right)
\end{align}
where $\varphi = \sqrt{h_i h_i + s^2}$, and $\{T^{\hat{a}}\}$ are the broken generators, spanning the coset $SO(6)/SO(5)$. 


After electroweak symmetry breaking, we can simplify the parameterisation by choosing $\pi_1 = \pi_2 = \pi_3 = 0, \pi_4 = h, \pi_5 = s$ in the unitary gauge. We can use the change of basis 
\begin{align}
h = \varphi \cos(\psi/f), && s = \varphi \sin(\psi/f)
\end{align}
to non-linearly recast the two physical fields $h,s$ into the fields $\psi, \varphi$. In the unitary gauge, the GB multiplet is
\begin{align}
\Phi_\text{unitary} = \left( 0,0,0,s_\varphi c_\psi,s_\varphi s_\psi, c_\varphi\right)
\label{unitary_sigma}
\end{align}
noting the shorthand $s_x = \sin\frac{x}{f},$ $c_x = \sin\frac{x}{f}$.

The GBs interact with the gauge sector through the covariant derivative

\begin{align}
\frac{f^2}{2}(D_\mu \Phi)^T (D^\mu \Phi) =& \frac{f^2}{2}\left[
\left( 0,0,0,\frac{\partial_\mu \varphi}{f} c_\varphi c_\psi - \frac{\partial_\mu \psi}{f} s_\varphi s_\psi,\frac{\partial_\mu \varphi}{f} c_\varphi s_\psi + \frac{\partial_\mu \psi}{f} s_\varphi c_\psi, -\frac{\partial_\mu \varphi}{f} s_\varphi\right)\right.\nonumber \\
& \left. - i g W^{a_L}_\mu T^{a_L} \Phi - i g' B_\mu T^{a_R} \Phi \right]^2\\
= & \frac{1}{2}(\partial \varphi)^2 + \frac{1}{2} (\partial \psi)^2 s_\varphi^2 + \frac{f^2}{8} s^2_\varphi c^2_\psi \left( g^2 W^2 + g'^2 B^2 + 2 g g' B_\mu W^{\mu,(3)}\right) \nonumber \\
&-(\frac{\partial_\mu \varphi}{f} c_\varphi c_\psi-	 \frac{\partial_\mu \psi}{f}s_\varphi s_\psi)\left(\frac{g}{2} W^{\mu,(3)} - \frac{g'}{2} B^\mu\right)s_\varphi c_\psi \label{eq:gauge_lagrangian} 
\end{align}
where $a_L$ runs from $1,2,3$, so $W^{\mu,(3)}$ is the third $W$ field. The third term in Equation \ref{eq:gauge_lagrangian} can be used to match to the SM Higgs-EW Lagrangian
\begin{align}
\mathcal{L}_{\text{Higgs-EW}} &= (D_\mu \Phi_{\text{SM}})^\dagger (D^\mu \Phi_{\text{SM}}) = (\partial H)^2 + \frac{1}{4}\left(v + H\right)^2 \left(2g^2 W_\mu^- W^{+\mu} + (g' B_\mu - g A_\mu^3)^2\right)  
\end{align}
We can then identify 
\begin{align}
v &= f\sin\frac{\langle\varphi\rangle}{f}\cos\frac{\langle\psi\rangle}{f}\label{eq:vev}
\end{align}
To make the match to the SM complete requires embedding these fields in $SU(2)_L \times SU(2)_R$ notation, and then we can redefine 
\begin{align}
W^2 \xrightarrow{SU(2)\times SU(2)} (W_L^1)^2 - (W_L^2)^2 + (W_L^3)^2 = 2W^+ W^- + (W_L^3)^2
\end{align} 
matching the SM coefficient. It is useful to define a ``vacuum misalignment'' - the degree to which the electroweak vacuum expectation value vector misaligns with the original $SO(6)$ vacuum expectation value:
\begin{align}
\xi \equiv \frac{v^2}{f^2} = \sin\frac{\langle\varphi\rangle}{f}\cos\frac{\langle\psi\rangle}{f}
\end{align}

\subsection{Matter content}

It is a well-known feature of CHMs that the gauge contribution to the NGB Higgs potential does not provide the correct sign for EWSB. Additionally, it is well-known that this potential contains divergent integrals unless some arbitrary cut-off is imposed, or some additional phenomenon regularises them. The solution to these problems is to include elementary and composite fermion sectors. They should both be embedded in some representation of $\mathcal{G}_0 = SO(6)$, and in this work we choose the fundamental representation. The embedding of the third generation quarks in {\bf 6} looks like:
\begin{align}
\psi_L &= \frac{1}{\sqrt{2}}\left(\begin{matrix}
b_L \\
-i b_L \\
t_L \\
it_L \\
0\\
0
\end{matrix} \right), &
\psi_R &= \left(\begin{matrix}
0\\
0\\
0\\
0\\
t_R e^{i\delta} \cos\theta\\
t_R \sin\theta
\end{matrix} \right)
\end{align}

%
%

The two sectors interact via mixing terms in the fermionic lagrangian, which is the most minimal set of interactions required to generate the SM Yukawas, in the unitary gauge (i.e. using the gauge symmetry to choose, $\langle h_i\rangle = 0, i=\{1,2,3\}$, giving $\Phi$ according to Equation \ref{unitary_sigma}): 

\begin{align}
\begin{split}
\mathcal{L}_f &= \bar{\psi}_L i \slashed{D} \psi_L + \bar{\psi}_R i \slashed{D} \psi_R + \Delta_{t_L} \bar{\psi}_L \Psi^T_R + \Delta_{t_R} \bar{\psi}_R\Psi^{\tilde{T}}_L \\
& + \bar{\Psi}^T_L(i\slashed{D} - m_T)\Psi^T_R + \bar{\Psi}^{\tilde{T}}_L(i\slashed{D} -m_{\tilde{T}})\Psi^{\tilde{T}}_R -Y_T\bar{\Psi}^T_L \Phi \Phi^\top \Psi^{\tilde{T}}_R - m_{Y_T}\bar{\Psi}^T_L \Psi^{\tilde{T}}_R + \textnormal{h.c.} \label{eq:fundamental_lagrangian}
\end{split}
\end{align}

Note the absence of terms $\bar{\Psi}^T_R  \Phi \Phi^\top \Psi^{\tilde{T}}_L$,  $\bar{\Psi}^T_L  \Phi \Phi^\top \Psi^T_R$ and $\bar{\Psi}^{\tilde{T}}_L  \Phi \Phi^\top \Psi^{\tilde{T}}_R$. We impose this absence in order to keep the Higgs potential finite. These terms could be introduced, in general, however the number of sites would then need to be extended from this minimal case, in order to cancel divergences in accordance with the Weinberg-like sum rules \cite{marzocca2012}. The elementary-composite mixing terms $\Delta_{t_L/t_R}$ have mass dimension one, as they contain the dynamics of the scalar link field. We now draw attention to features that differentiate this model from the MCHM. These include two elementary embeddings of the partially composite top quark. Under $SO(4)\approx SU(2)_L \times SU(2)_R$, we have the decomposition $(\bm{2,2}) \oplus (\bm{1,1}) \oplus (\bm{1,1})$. The left-handed top quark is embedded into the $(\bm{2,2})$, which protects the $Zb_L \bar{b}_L$ coupling, whilst the right-handed top quark is embedded as a linear combination in both singlets. $\delta$ appears due to a choice of this top coupling. It is not a physical parameter, and can be removed by a phase transformation under the $SO(2)$ subgroup of $SO(6)$, taking $\me^{i\delta} \rightarrow i$. $\theta$ is however an important artefact of the NMCHM, and appears from the choice of composite partner embedding within the $SO(2)$ subgroup. As in previous work, we include only the top quark and heaviest quark doublet in the analysis. We do not consider partially composite leptons in this study, such as were previously studied in the context of the MCHM in \cite{carmona2015, Barnard:2017kbb}.

The elementary terms\footnote{After expanding the 6-plets, one can group the left-handed terms $\{t_L,b_L\}$ into their regular SM doublet $q_L$.} appear in an effective Lagrangian, coming from decomposing the GB and fermion multiplets in the unitary gauge under $SU(2) \times SU(2)$, given in \cite{Redi:2012ha} 
\begin{align}
\begin{split}
\mathcal{L}_\textnormal{fermion} &= \bar{q}_L\Pi_q (p^2, \psi, \varphi)  \slashed{p} q_L + \bar{t}_R\Pi_t (p^2, \psi, \varphi) \slashed{p} t_R +\bar{q}_L M_t(p^2, \psi, \varphi) t_R + \textnormal{h.c} \\
&= \bar{q}_L \left( \frac{\Delta^2}{(\Delta_{t_L})^2} + \Pi_{q_L,0}(p^2) + \frac{1}{2} \frac{s^2_\varphi}{\varphi^2}\Pi_{q_L,1}(p^2)H^c H^c\right) \slashed{p} q_L \\
&+ \bar{t}_R \left( \frac{\Delta^2}{(\Delta_{t_R})^2} + \Pi_{u_R,0}(p^2) + s^2_\theta \Pi_{u_R,1}(p^2) + \left[s_\varphi^2 (c_\theta^2 s_\psi^2 - s_\theta^2)\right] \Pi_{u_R,1}(p^2) \right) \slashed{p} t_R \\
&+\bar{q}_L \frac{M_{u,1}}{\sqrt{2}}\frac{s_\varphi}{\varphi}H^c \left( ic_\theta s_\varphi s_\psi + s_\theta c_\varphi\right) t_R + \textnormal{h.c}
\end{split}\label{eq:fermion_lagrangian}
\end{align}
where $q_L$ and $H^c$ are the SM quark doublet and charge conjugate of the normalised SM Higgs doublet, respectively,
\begin{align}
q_L = \left( t_L , b_L \right)^T,  && H^c = \frac{1}{h} i\sigma_2 \left( \begin{matrix}
h^1 - ih^2\\
h^3 - ih^4
\end{matrix}\right)^* = \frac{1}{h} \left( \begin{matrix}
-(h^1 + ih^2)\\
h^3 + ih^4
\end{matrix}\right)
\end{align}
The form factors $\Pi_i$ are given in full in Appendix \ref{expressions}. Here, the elementary bare quark mass terms $\frac{\Delta^2}{(\Delta_{t_L / t_R})^2}$ can be understood as canonically normalised. That is, there is some common scale $\Delta$ that can be factored out once the form factors are found.

%

A final distinction of the NMCHM is the relevance of only one choice of representation (although many composite partners and resonances could be added in this representation)~\cite{gripaios2009beyond}. In brief, the three smallest representations under $SO(6) \approx SU(4)$ are the $\textbf{4}$, the $\textbf{6}$, and the $\textbf{10}$. The $\textbf{4}$ does not contain a bidoublet when decomposed under $SU(2)_L \times SU(2)_R$, and thus cannot contain a representation that couples with the SM quark doublet, which must also be incompletely embedded into a $\textbf{4}$. The symmetric traceless $\textbf{10}$ does contain such a bidoublet, however upon embedding the SM quarks in a simple way, we see that there remains a $U(1)_s$ symmetry protecting the scalar singlet. In this case, the singlet will correspond to an electroweak axion, with properties that have been excluded experimentally. Less minimal \textbf{10} embeddings have been shown in Reference \cite{Serra:2015xfa} to produce a massive singlet and evade exclusion. Thus, this leaves the $\textbf{6}$ as the simplest representation for the quark partners, and we thus focus on the NMCHM$^\textbf{6}$.

\subsection{Goldstone Boson Vacuum Behaviour}\label{sec:GB_Vacuum}

After EWSB, we can write the low energy effective potential for the interactions of the Higgs boson and scalar singlet with gauge fields and fermions \cite{contino2010strong} 
\begin{align}
\begin{split}
\mathcal{L}_\text{eff} &= \frac{1}{2}(\partial_\mu h)^2 + \frac{1}{2}(\partial_\mu s)^2 - V(h,s) + \frac{v^2}{4}\textnormal{Tr} \left[ D_\mu \Sigma^\dagger D^\mu \Sigma \right] \left( 1 + 2a_h \frac{h}{v} + b_h \frac{h^2}{v^2} + b_s \frac{s^2}{v^2} + ... \right) \\
- & m_i \bar{\psi}_{Li} \Sigma\left(1 + c_h \frac{h}{v} + ...\right) \psi_{Ri} - m_i \bar{\psi}_{Li} \left(c_s \frac{s}{v} + ... \right) \psi_{Ri} + h.c. \label{eq:general_lagrangian}
\end{split}
\end{align}

where the GBs eaten by the $W$ and $Z$ bosons are parameterised by $\Sigma=\exp(i\chi^a\sigma^a/v)$. The couplings of $a$, $b$ and $c$ can be obtained as:
\begin{align}
a_h = \sqrt{1-\xi}, && b_h = 1-2\xi, && b_s = 1, && c_h = \frac{1-2\xi}{\sqrt{1-\xi}}, && c_s = i\frac{\xi}{1-\xi}\cot\theta
\end{align}

To compute the vacuum misalignment $\xi$ and therefore the coupling terms, we need to explore the effective potential of the Goldstone bosons. This is generically given by the Coleman-Weinberg formula for the gauge boson and top quark contributions, where the form factors are given in appendix \ref{expressions}
\begin{align}
V_\textnormal{fermionic} &= \frac{9}{2}\int\frac{d^4p}{(2\pi)^2}\log\Pi_W - 2N_c \int \frac{dp^4}{(2\pi)^4} \ln \left( p^2 \Pi_{t_L} \Pi_{t_R} - \Pi_{t_L t_R}^2\right)\label{eq:fermionic_potential}
\end{align}

As in the MCHM, we require this potential to have a minimum such that it reproduces the electroweak vacuum expectation value (VEV). We can attempt to do this at leading order, which would lead to a natural EWSB potential. For example, the potential in Equation \ref{eq:fermionic_potential} can be expanded at leading order in the MCHM Goldstone field as 
\begin{align}
V(h) = \alpha \sin^2\frac{h}{f} + \mathcal{O}(s_h^4)
\end{align}
This has possible minima\footnote{Depending on the sign of $\alpha$} at integer multiples of $\langle h \rangle = \frac{f\pi}{2}$, which is far too high. The case of $\langle h \rangle = 0$ leads to no EWSB. The same obstacle applies to the NMCHM potential, which at leading order in $\varphi, \psi$ is
\begin{align}
V(\varphi, \psi) = \sin^2\frac{\varphi}{f}\left( c_1 + c_2 \sin^2\theta - c_3 \sin^2\theta \right) + \mathcal{O}(s_\varphi^4,s_\psi^4)
\end{align}
where the expressions for the integral terms $c_i$ are given in Appendix \ref{expressions}. This has stationary points at integer multiples of $\langle \varphi \rangle = \frac{f\pi}{2}$. 
Again, this is problematic, as we need the EW VEV $v =  f \sin\frac{\langle\varphi\rangle}{f}\cos\frac{\langle\psi\rangle}{f}$ to be at a much lower scale than the typical symmetry breaking scale $f=f \sin\frac{\pi}{2} \cos{0}$.

Therefore, as in the MCHM, we must break EW symmetry by considering higher-order terms that must cancel precisely, requiring the notorious composite Higgs double tuning. We include higher order terms, up to quartic in $\sin\frac{\varphi}{f}\sin\frac{\psi}{f}$ \footnote{We include such seemingly high order terms since $\xi^2 \propto s_{\langle \varphi \rangle}^2 s_{\langle \psi \rangle}^2$, and must therefore include each field up to consistent order. Note that to obtain Equations \ref{eq:varphi} and \ref{eq:higgs_mass}, it is sufficient to expand to quadratic order $V = s_\varphi^2(c_1 + c_2 s_\theta^2 - c_3 s_\theta^2) - s_\varphi^2 s_\psi^2(c_1 + c_2 c_\theta^2 -c_3 s_\theta^2) + s_\varphi^4 c_3 s_\theta^2$. The singlet mass, on the other hand, requires corrections given by the quartic-order potential.}
\begin{align}
\begin{split}
V(\psi,\varphi) = c_1 \sin^2\frac{\varphi}{f}\cos^2\frac{\psi}{f} + c_2 \sin^2 \frac{\varphi}{f} \left( \sin^2 \theta - \cos^2 \theta \sin^2 \frac{\psi}{f} \right) \\
- c_3 \sin^2 \frac{\varphi}{f} \cos^2 \frac{\psi}{f} \left( \cos^2 \theta \sin^2 \frac{\varphi}{f} \sin^2 \frac{\psi}{f} + \sin^2 \theta \cos^2 \frac{\varphi}{f}\right)\\
 + \mathcal{O}(\sin^{10}_\varphi, \sin^8_\varphi\sin^2_\psi,...,\sin^{10}_\psi)
\end{split}
\label{potential}
\end{align}
To find the classical expectation value, we solve for
\begin{align}
\frac{\partial V}{\partial \psi} \bigm\lvert_{\varphi = \langle \varphi \rangle, \psi = \langle \psi \rangle} & = 0\\
\frac{\partial V}{\partial \varphi} \bigm\lvert_{\varphi = \langle \varphi \rangle, \psi = \langle \psi \rangle} & = 0
\end{align}
where zeroes will be found both from trivial extrema (i.e. integer multiples of $\varphi, \psi = \frac{f\pi}{2}$) and double tuning extrema (cancellations between terms, requiring tuning of $c_1, c_2, c_3$). It can be shown that the surface $\langle\psi\rangle =0$ or $\langle \varphi \rangle = \frac{f\pi}{2}$ always contains an extremum of the potential, and either (but not both) can be chosen such that EWSB may still occur realistically. 

\begin{figure}
\begin{subfigure}{0.5\textwidth}
\includegraphics[scale=0.32]{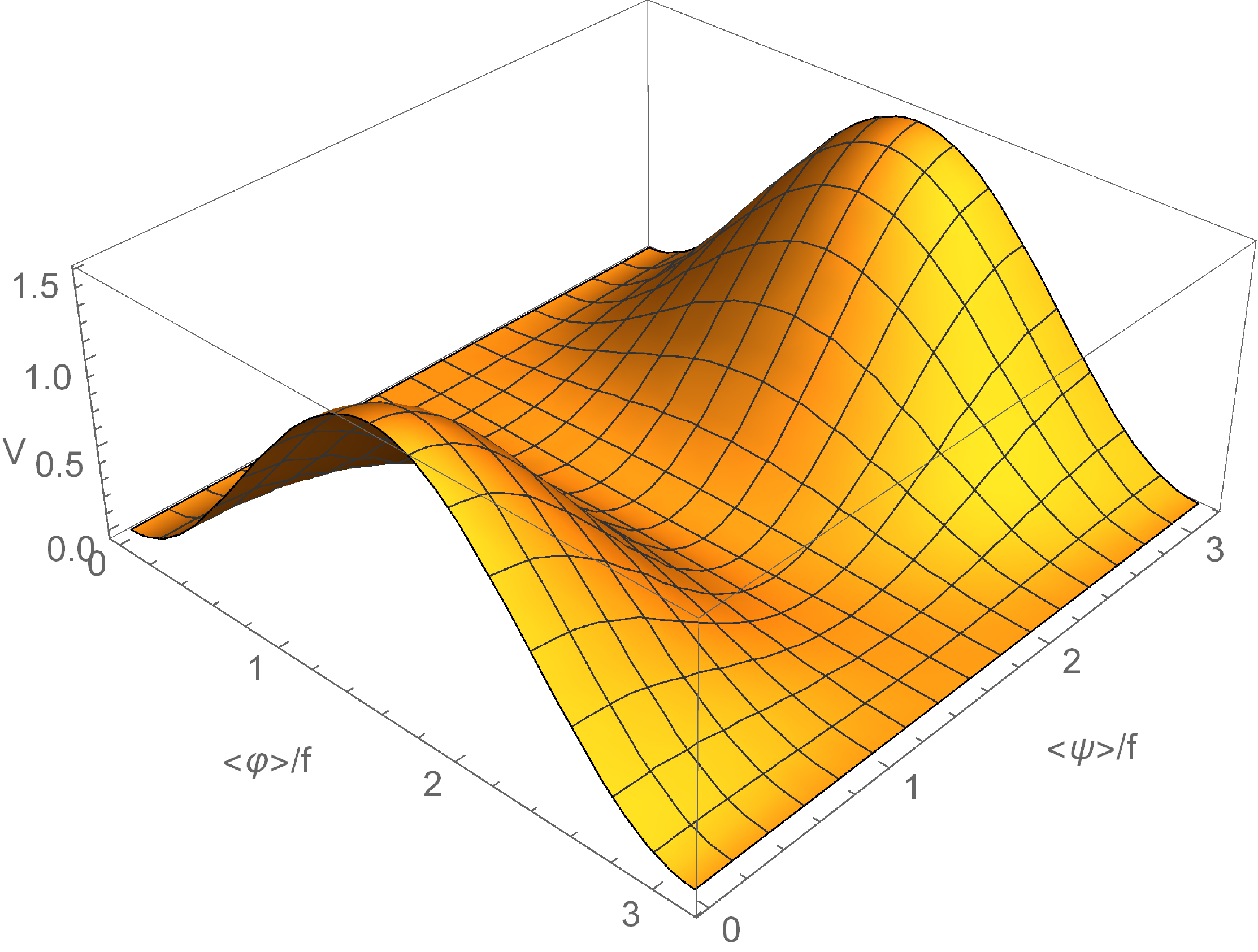}
\end{subfigure}%
\begin{subfigure}{0.5\textwidth}
\includegraphics[scale=0.32]{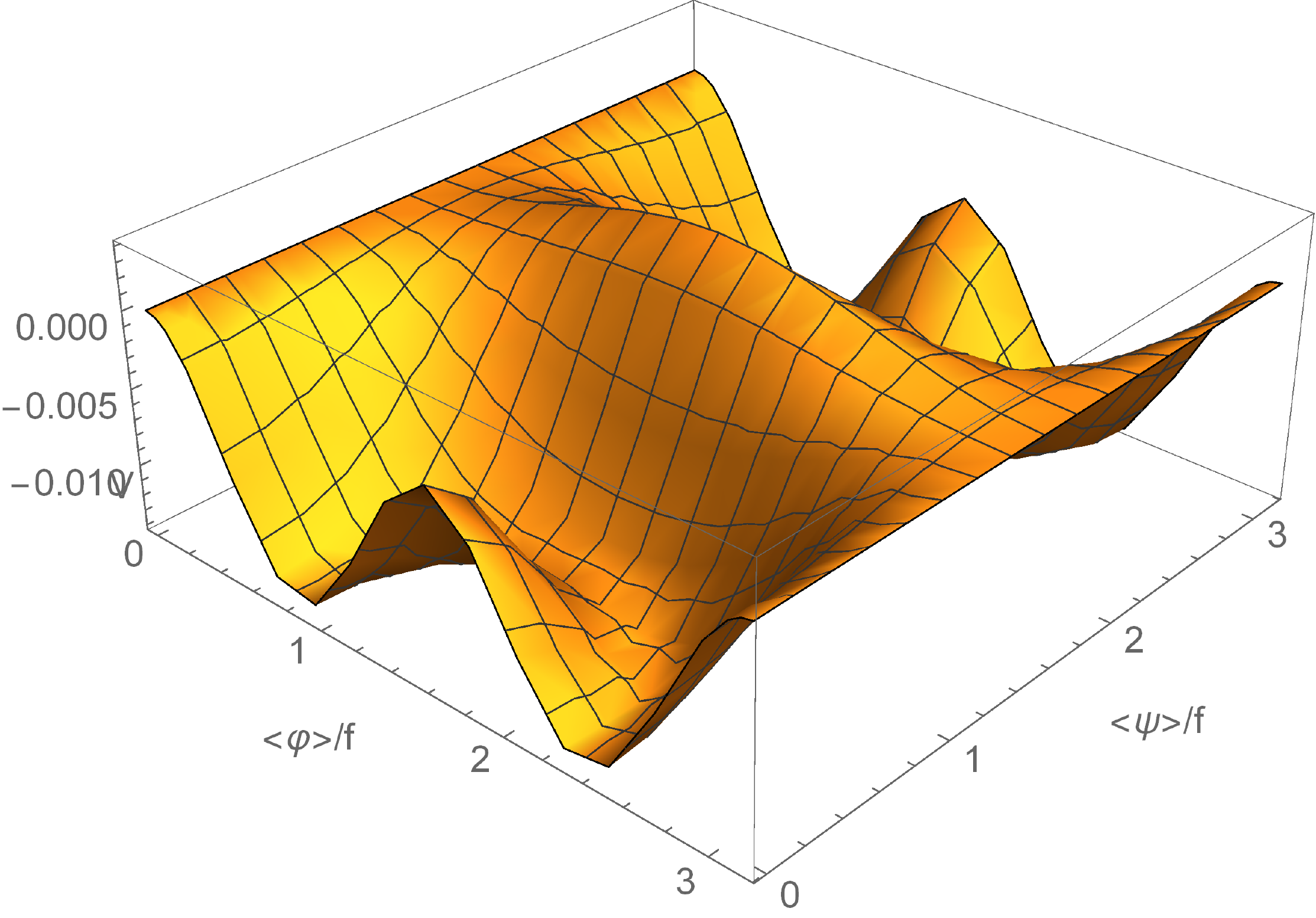}
\end{subfigure}
\caption{Two examples of the GB potential. On the left, $c_1 = 1, c_2 = 1, c_3 = -0.1, s_\theta = 0.7$ with $\xi = 15.7$, corresponding to no EWSB. On the right, $c_1 = 0.1, c_2 = -0.2, c_3 = 0.1, s_\theta = 0.7$, with $\xi = 0.48$. Satisfying the condition $\xi<1$ allows for the possibility of EWSB.}\label{fig:potential}
\end{figure}

We take $\langle\psi\rangle =0$ to give a stationary point and thus give the singlet no VEV. See Reference~\cite{Redi:2012ha} for a discussion of the validity of this choice. For this choice, a potential extremum is found for
\begin{align}
\sin\frac{\langle\varphi\rangle}{f} &= \sqrt{\frac{c_3 s^2_\theta - c_1 - c_2 s^2_\theta }{2 c_3 s^2_\theta}}\label{eq:varphi}\\
\implies \xi &= \frac{v^2}{f^2} = \frac{c_3 s^2_\theta - c_1 - c_2 s^2_\theta }{2 c_3 s^2_\theta}
\end{align}
using the definition in Equation~\ref{eq:vev}. This implies that  $0 < \frac{c_3 s_\theta^2 - c_1 - c_2 s_\theta^2}{2c_3 s_\theta^2} < 1$, in order to achieve a non-trivial VEV. This can be used as a constraint to rescale $f$ for correct EWSB behaviour\footnote{Note that this constraint is not sufficient for EWSB - it only corresponds to an extremum. The Higgs mass must be found to be positive, to ensure that this solution is a local minimum.}. To better illustrate the possible behaviour of the potential, we show it in Figure \ref{fig:potential} for two different sets of \{$c_1,c_2,c_3,s_\theta$\}. The first plot shows the typical case encountered in much of the parameter space where the extrema are given only by integer multiples of $\varphi, \psi = \frac{f\pi}{2}$, leading to no EWSB. The second plot shows an example of the fine cancellations which occur in a small region of the parameter space, corresponding to the solution in Equation \ref{eq:varphi}. This gives additional minima and maxima, which are a condition of EWSB.

The masses of the two scalars can be found using the second derivatives of Equation \ref{potential}, and the solution for $\langle \varphi \rangle$ in Equation \ref{eq:varphi}
\begin{align}
m_\varphi^2 = m^2_h &= \frac{-4c_1 c_2 -2c_1^2 / s_\theta^2 + 2(c_3^2 - c_2^2)s_\theta^2}{c_3 f^2}\label{eq:higgs_mass}\\
m^2_\psi = m^2_s &= \frac{c_1 - (c_2 + c_3)s_\theta^2}{s_\theta^2 f^2} c_{2\theta}\label{eq:singlet_mass}
\end{align}
Note that we have changed the basis from $\varphi, \psi$ to $h,s$, but that the masses are the same due to $\langle\psi\rangle = 0$ being a stationary point. This can be laboriously shown with liberal application of the chain rule.

We can thus analyse the Higgs mass expression as a function of each of the integral terms. This also gives our top mass term, which can be found by diagonalising the low-energy Lagrangian in Equation~\ref{eq:fermion_lagrangian} 
\begin{align}
\abs*{m_t}^2 &= \frac{[M_1^u(0)]^2}{\Pi_{t_L}(0) \Pi_{t_R}(0)}s_{\langle\varphi\rangle}^2 c_{\langle\psi\rangle}^2 \left( c_\theta^2 s_{\langle\varphi\rangle}^2 s_{\langle\psi\rangle}^2 +s_\theta^2 c_{\langle\varphi\rangle}^2 \right)\\
&= \frac{[M_1^u(0)]^2}{\Pi_{t_L}(0) \Pi_{t_R}(0)} s_\theta^2 \xi (1-\xi)\label{eq:top_mass}
\end{align}

\section{Scan details}
\label{scanning}

The NMCHM as parameterised in Equations \ref{eq:fundamental_lagrangian} and \ref{eq:fermion_lagrangian} (with the correlators given in the appendix), contains the following 10 independent parameters:

\begin{itemize}
\item The bare masses of the lightest scalar resonances $m_\rho, m_a \in [0.3,10]$ TeV; 
\item The ratio of composite-elementary mixing in the gauge sector $t_\theta = \frac{g_2}{g_\rho}\in[0,1]$; 
\item The on-diagonal bare masses of the top partners $m_{T},m_{\tilde{T}} \in [0.3,10]$ TeV;
\item The off-diagonal bare mass of the top partners $m_{Y_T} \in [0.3,10]$ TeV; 
\item The proto-Yukawa couplings $Y_T \in [-10,10]$ TeV;
\item The extent to which the observed SM quark doublet and top singlet are composite $d_Q, d_T = \Delta_{q,t}/m_{T,\tilde{T}}\in [0,3]$;
\item The top quark eigenstate angle in the $SO(2)$ subgroup $\theta \in [0,\frac{\pi}{2}]$;
\end{itemize}

In order to produce a well-sampled analysis of the model's fine tuning, we use the {\tt Diver} implementation of the differential evolution algorithm to find physical regions of the model's parameter space~\cite{Workgroup:2017htr,storn1997differential}. This has proved particularly useful in finding optimum regions in difficult likelihood functions, such as those encountered in Higgs portal dark matter and supersymmetric examples~\cite{Athron:2017qdc,Athron:2017yua,Cornell:2016gho}.

The algorithm first randomly seeds the parameter space with a \emph{population} of $NP$ vectors $\{\textbf{X}_i^g\}$, where $i$ indexes the members of the population, and $g$ indexes the generation. Subsequent generations of the population are then obtained by performing mutation, crossover and selection steps, and these are repeated at each future generation.

The \emph{mutation} step produces a set of \emph{donor vectors} $\{\textbf{V}_i\}$ from the current population of vectors $\{\textbf{X}_i^g\}$. The production of each donor vector $\textbf{V}_i$ occurs by choosing three random vectors $\textbf{X}_{r1}$, $\textbf{X}_{r2}$ and $\textbf{X}_{r3}$ from the current population (on the condition that none of these are the same, and that none of them matches $\textbf{V}_i$). $\textbf{V}_i$ is then taken to be:
\begin{equation}
\textbf{V}_i=\textbf{X}_{r1}+F(\textbf{X}_{r2}-\textbf{X}_{r3})
\end{equation}
where $F$ is a parameter that controls the strength of the differential variation.

The \emph{crossover} step is then used to produce a set of trial vectors $\{\textbf{U}_i\}$ that will potentially form the next generation of vectors. For the $k$th component of the trial vector $\textbf{U}_i$, a random number between 0 and 1 is chosen. If this number is less than or equal to a parameter $Cr$ (chosen in advance of the scan), then the component is taken from the corresponding donor vector $\textbf{V}_i$. Otherwise, the component is taken from the corresponding vector in the previous generation. After all of the components of $\textbf{U}_i$ have been chosen, one component is reassigned, thus ensuring that the trial vectors and their corresponding vectors in the current generation are always different. A component $l$ of the vector is chosen at random, and the trial vector component is set to the donor vector value, irrespective of its previous value.

Finally, a \emph{selection} step is used to choose the vectors for the next generation. The value of the likelihood function for each vector in the current generation $\textbf{X}_i^j$ is compared with the likelihood for the correspondng trial vector $\textbf{U}_i$, and the points with higher likelihood are retained for the next generation.


We use a multivariate Gaussian likelihood function that takes as inputs three values. The first two are physical observables we wish to reproduce: the masses of the SM Higgs $m_h$ and top quark $m_t$. The particular values for the observables $\mathcal{O}_i$ used in this scan were $\mathcal{O}_1 = m_h^{\text{exp}} = 125\pm 1$ GeV and $\mathcal{O}_2 = m_t^{\text{exp}} = 155\pm 1$ GeV; where the uncertainties are not chosen to reflect the known experimental uncertainties, instead being used to control how precisely the central values are reproduced by our scanning method \footnote{The values are not precisely the experimentally determined values. They have strong and electroweak RGE running applied, as outlined in \cite{xing2007}.}. The third value is the measure of higher-order tuning $\Delta$ defined in the next section. We are interested in exploring areas of low tuning, and so penalise parameter points with a function of $\Delta$. The cost function (which also defines the likelihood $L$) is then
\begin{align}
L = \exp\left(-\frac{(125 - m_h)^2}{2} - \frac{(155 - m_t)^2}{2} - \frac{\Delta^2}{2\times 1000^2}\right)
\end{align} 
The cost of tuning is heavily scaled down, since the tuning has a minimum at $\mathcal{O}(10)$, and this cost factor would dominate the scan otherwise.

This {\tt Diver} package optimises the differential evolution algorithm further by allowing $Cr$ and $F$ to evolve, called Adaptive Differential Evolution. This occurs in the intuitive way - by sampling $Cr$ and $F$ uniformly in the seeding step, and subsequently propagating those values that lead to lower cost function outputs. We enabled this adaptivity, and in doing so found a suitable set of parameter points (i.e., giving valid EWSB, with SM masses within two $\sigma$ of the measured values) significantly faster than with other scanning techniques (based on our previous experience with Markov Chain Mote Carlo algorithms and nested sampling). 

In the following section, we choose to study the subset of points that are in the vicinity of the correct SM behaviour by applying observable cuts as follows:
\begin{align}
\{120,140,800\}\; \gev \leq \{m_h, m_t, f\} \leq \{130,170, \infty \}\; \gev
\end{align}
and we also require all parameters with mass dimension to be less than $4\pi f$, to be within the perturbative limit.
We then calculate the spectrum of resonances and the expected deviations from the SM Higgs couplings. The latter are parameterised as a fraction of the composite Higgs-$\chi$-$\chi$ coupling $c$ (where $\chi$ is any of the SM states that the Higgs can couple to) with the SM Higgs-$\chi$-$\chi$ coupling $c_\textnormal{SM}$,
\begin{align}
r_\chi &= \frac{c(h\chi\chi)}{c_\textnormal{SM}(h\chi\chi)}\label{coupling}\, .
\end{align}
%



\section{Fine-Tuning}
\label{results}
\subsection{Fine-tuning measure}
To calculate the fine-tuning of our parameter points, we use a more accurate measure than the usual Barbieri-Giudice (BG) measure. This concept was developed in Reference~\cite{Barnard:2015ryq}, and further generalised in Reference~\cite{Barnard:2017kbb}, and we here provide a brief summary\footnote{See \cite{Fichet:2012sn} for a derivation of the measure from Bayesian reasoning}.

Consider the usual BG measure
\begin{align}
\Delta_{BG} = \max_{i,a}\abs*{\frac{x_i}{\mathcal{O}_a}\frac{\partial \mathcal{O}_a}{\partial x_i}}_{\mathcal{O}=\mathcal{O}_\textnormal{exp}}.
\end{align}
That is, the maximum tuning over each observable $\mathcal{O}_a$, with respect to each parameter $x_i$, evaluated at the experimental values. While a useful heuristic, this measure does not appropriately punish models that decrease their fine-tuning by increasing the number of parameters. To account for this, one can simply treat the tuning for each observable as a vector,
\begin{align}
\nabla^\mathcal{O} = \frac{x_i}{\mathcal{O}}\frac{\partial \mathcal{O}}{\partial x_i}_{\mathcal{O}=\mathcal{O}_\textnormal{exp}}\label{BG}\, ,
\end{align}
before defining an overall first-order of tuning as the average over the magnitudes of these tuning vectors,
\begin{align}
\Delta_1^{\mathcal{O}_a} = \abs*{\nabla^{\mathcal{O}_a}} && \implies && \Delta_1 = \frac{1}{n_\mathcal{O}}\sum\limits_{a=1}^{n_\mathcal{O}} \abs*{\nabla^{\mathcal{O}_a}}\label{fot}
\end{align}

However, one can see that this still doesn't account for the often complex interdependencies between parameters or observables, e.g. a Higgs mass and top quark mass that may depend on some common parameters. A new measure can account for this higher order tuning using the determinant of the set of observable vectors
\begin{align}
\Delta_2^{ab} = \begin{vmatrix}
\nabla^{\mathcal{O}_a}\cdot\nabla^{\mathcal{O}_a} & \nabla^{\mathcal{O}_a}\cdot \nabla^{\mathcal{O}_b}\\
\nabla^{\mathcal{O}_a}\cdot\nabla^{\mathcal{O}_b} & \nabla^{\mathcal{O}_b}\cdot \nabla^{\mathcal{O}_b}
\end{vmatrix}^\frac{1}{2}_{\mathcal{O}=\mathcal{O}_\textnormal{exp}\, .}\label{double}
\end{align}
and these sum up to give a full ``double" tuning,
\begin{align}
\Delta_2 = \frac{1}{2}(\Delta_2^{ab} +\Delta_2^{bc}+\Delta_2^{ca})\, .\label{totaldouble}
\end{align}
This generalises to more than three observables in a straight-forward way (see Reference~\cite{Barnard:2017kbb} for details). The full tuning $\Delta$ is then the sum of all orders of tuning,
\begin{align}
\Delta = \sum\limits_{a=1}^{n_\mathcal{O}} \Delta_a.  \label{hot}
\end{align}


\subsection{Fine-tuning results}
We now present the scan results in terms of the fine-tuning found at each viable parameter point. The tuning of each point is shown against the lightest vector-boson resonance mass $m_\rho$, the lightest top partner resonance mass, the mass of the $SO(6)$ scalar singlet, the Higgs coupling ratios $r_\chi$ and the vacuum misalignment $\xi = v^2/f^2$. A convex hull is provided to understand the general limits of minimal fine-tuning (note that given the logarithmic scale, the hull may not always appear to be convex). In all coupling correction plots, several predicted bounds are included, based on the anticipated precision of the future International Linear Collider (ILC)~\cite{tian2016measurement}. Two bounds are included - a pessimistic bound at the $250$GeV baseline ILC, and an optimistic bound from a high-energy, high-luminosity upgrade. These bounds are given in Table \ref{tab:ILC}.

\begin{figure}[h]
\centering
\begin{subfigure}{0.5\textwidth}
\centering
\includegraphics[width=1\linewidth]{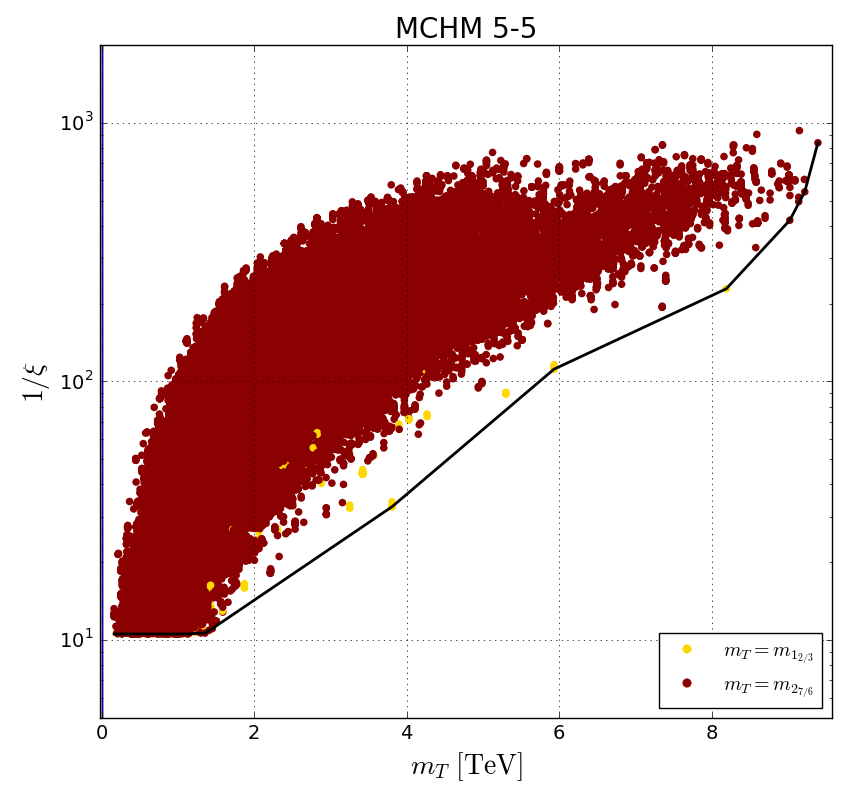}
\end{subfigure}%
\begin{subfigure}{0.5\textwidth}
\centering
\includegraphics[width=1\linewidth]{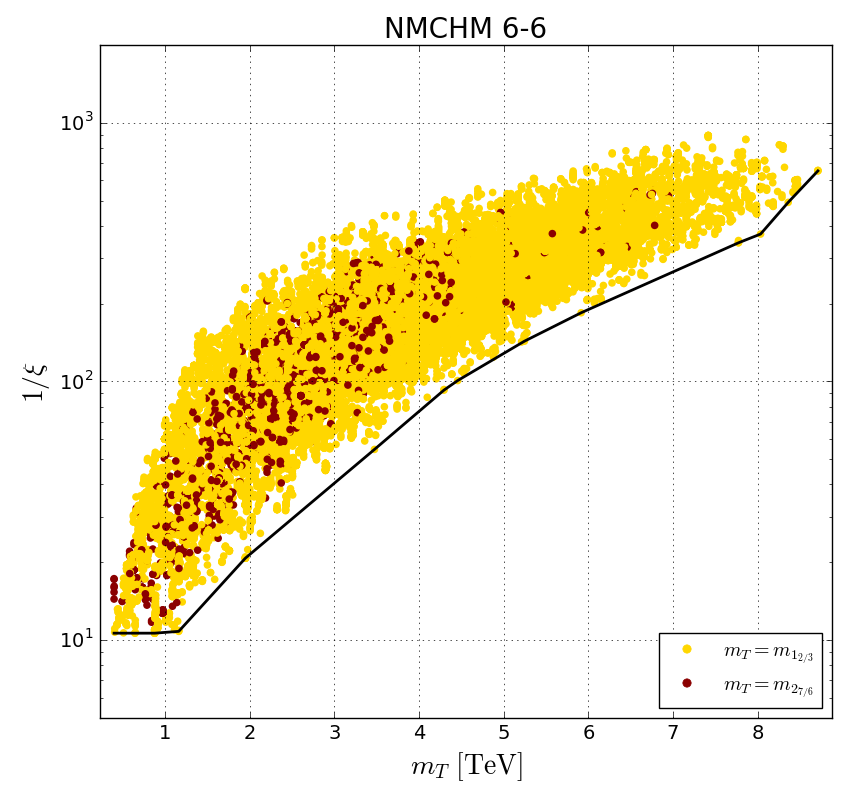}
\end{subfigure}\caption{Comparison of each model's lightest top partner vs. naive tuning}\label{fig:masses_vs_xi}
\end{figure}

Before analysing our results, we note that an earlier study (Reference~\cite{Banerjee:2017qod}) demonstrated that higher top partner masses may be achieved in the NMCHM, with no fine-tuning penalty, through a process dubbed ``level repulsion''. If the doublet and singlet in the pNGB sector both get VEVs, the model exhibits a tree-level doublet-singlet mixing. If the singlet state is heavier, then the mixing can result in pushing down the dominantly  doublet eigenstate to match the observed Higgs mass at 125 GeV. Before mixing, the masses of both of the states can conceivably be larger, which makes the theory more natural. This earlier result may naively appear to conflict with the results of the previous section, but in fact there is no contradiction once one compares the different scope of the studies and the fine-tuning measure used. The requirement that both the doublet and singlet get a VEV corresponds to $\theta$ being close to $\pi/2$, and thus this is a special limit of the more general theory (one that would in fact appear as a fine-tuning contribution in a proper analysis). We assume in our study that the singlet does not acquire a VEV, meaning that there is no overlap between our results and the previous study. Indeed, if we examine the naive tuning measure of $1/\xi$ as a function of the lightest top partner (LTP) mass in our study, we find no tuning gain for the NMCHM vs the MCHM (see Figure~\ref{fig:masses_vs_xi}).

\begin{figure}[h]
\centering
\begin{subfigure}{0.5\textwidth}
\centering
\includegraphics[width=1\linewidth]{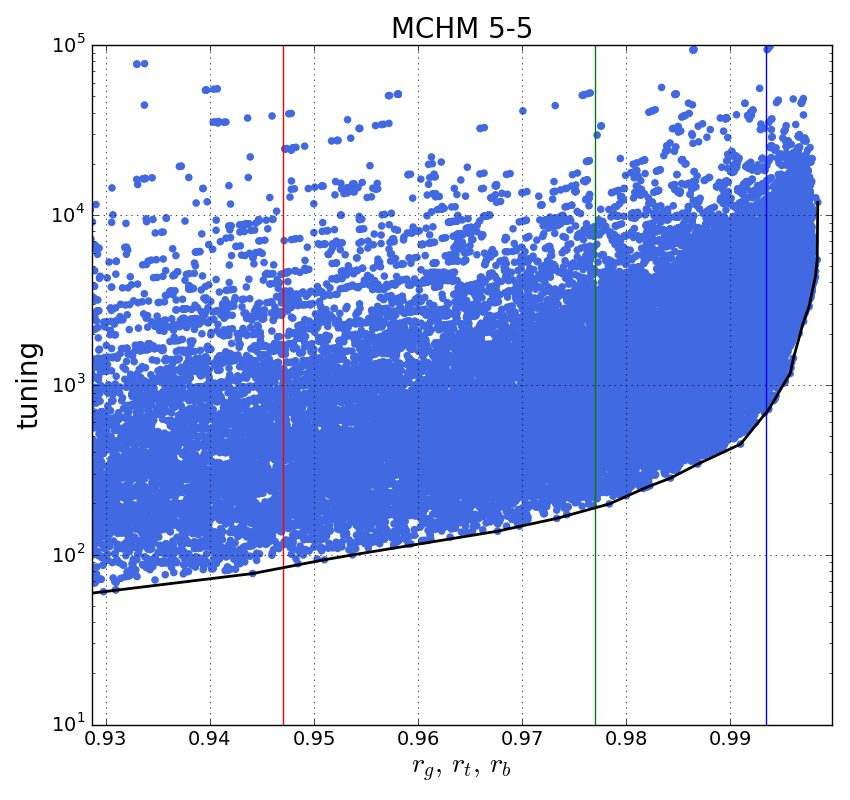}
\end{subfigure}%
\begin{subfigure}{0.5\textwidth}
\centering
\includegraphics[width=1\linewidth]{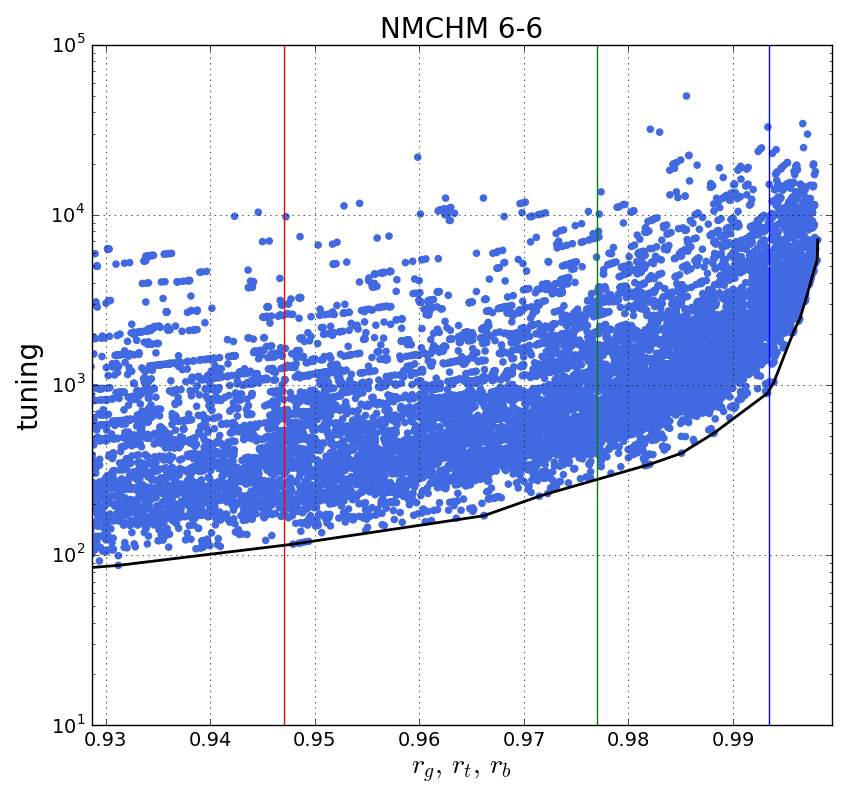}
\end{subfigure}\caption{\label{fig:tuning-vs-couplings} Comparison of higher-order tuning (defined in Equation \ref{hot}) in the Higgs-gluon, -top and -bottom coupling deviation (as defined in Equation \ref{coupling}) between the minimal and next-to-minimal models. Precision bounds (denoted by coloured lines) are defined in Table \ref{tab:ILC}. The red line shows the expected precision of a $250$ GeV ILC, green a $500$ GeV ILC, and blue a high-luminosity $1$ TeV ILC.}
\end{figure}

In Figure~\ref{fig:tuning-vs-couplings} we show the higher-order tuning as a function of the modification to the Higgs-gluon, Higgs-top and Higgs-bottom couplings, for both the MCHM and NMCHM models. As one would expect, the minimum fine-tuning available in each model would increase if one were able to measure the Higgs couplings more precisely (assuming that they remain at the SM values). In both models, the impact of a 250 GeV ILC is minimal, but the high-luminosity 1 TeV ILC would increase the minimum fine-tuning by roughly an order of magnitude. We also observe a slightly higher fine-tuning in the NMCHM model, relative to the MCHM model, regardless of future measurements of the Higgs couplings. This can be attributed to a small punishment for increasing the parameter set from nine to ten. A thorough discussion of parameter set scaling in the higher order tuning measure can be found in reference \cite{Barnard:2017kbb}. This agrees with a first-order expectation, since NMCHM observables are generically proportional to MCHM observables according to $m_\text{NMCHM} \propto m_\text{MCHM}\sin\theta$, and $\theta$ is a free parameter.

\begin{figure}[h]
\centering
\begin{subfigure}{0.5\textwidth}
\centering
\includegraphics[width=1\linewidth]{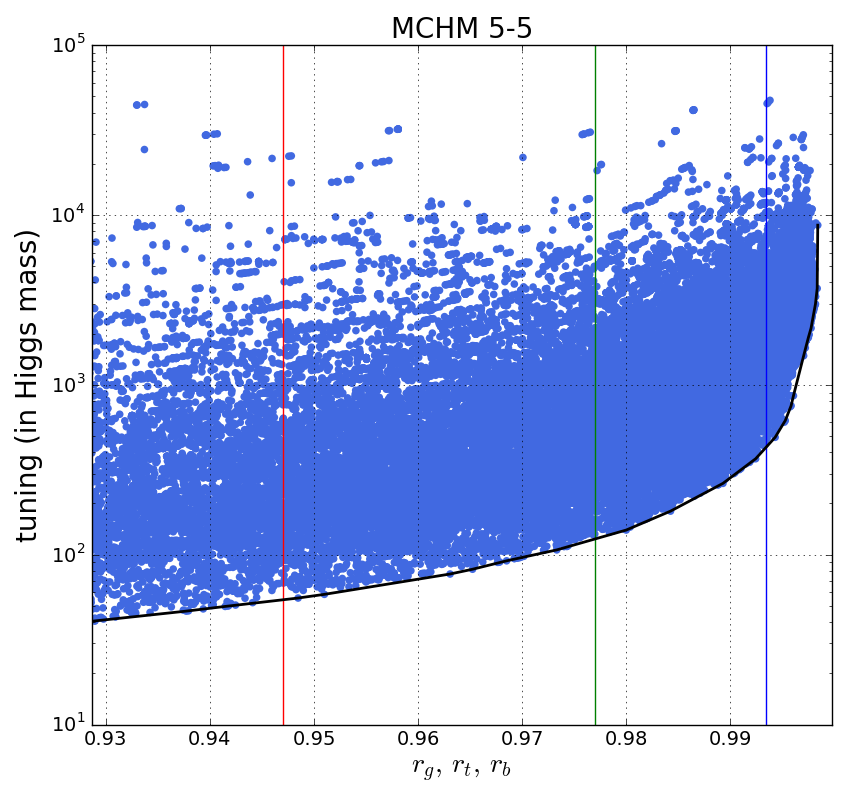}
\end{subfigure}%
\begin{subfigure}{0.5\textwidth}
\centering
\includegraphics[width=1\linewidth]{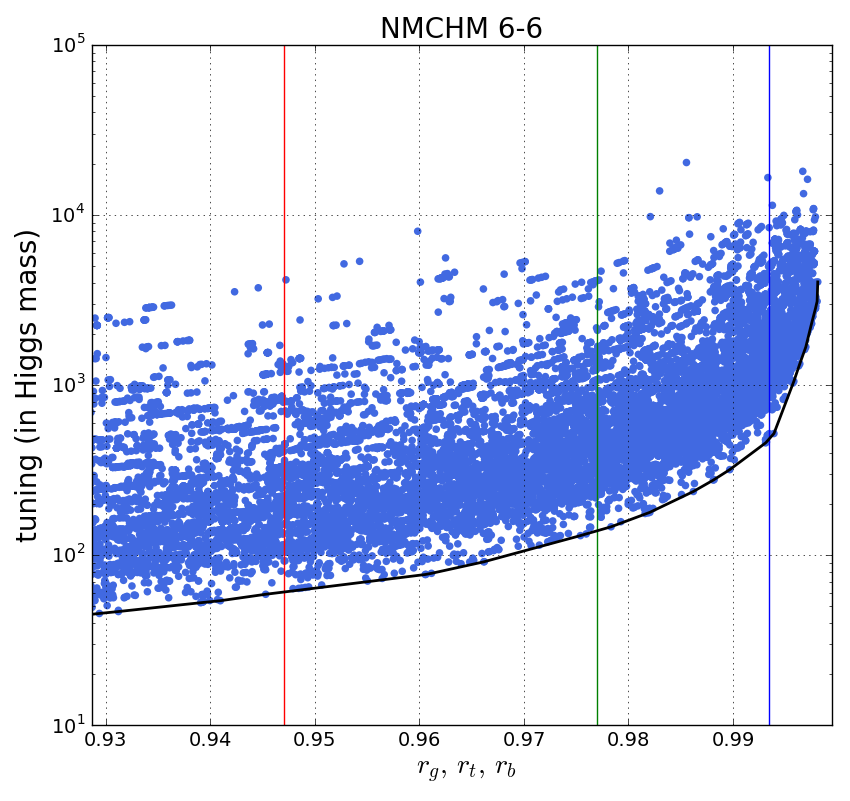}
\end{subfigure}\caption{\label{fig:first-tuning-higgs-vs-couplings} Comparison of the first-order tuning (as defined in Equation~\ref{fot}) contribution from the Higgs mass, in the Higgs-gluon, -top and -bottom coupling deviation. The red line shows the expected precision of a $250$ GeV ILC, green a $500$ GeV ILC, and blue a high-luminosity $1$ TeV ILC.}
\end{figure}

\begin{figure}[h]
\centering
\begin{subfigure}{0.5\textwidth}
\centering
\includegraphics[width=1\linewidth]{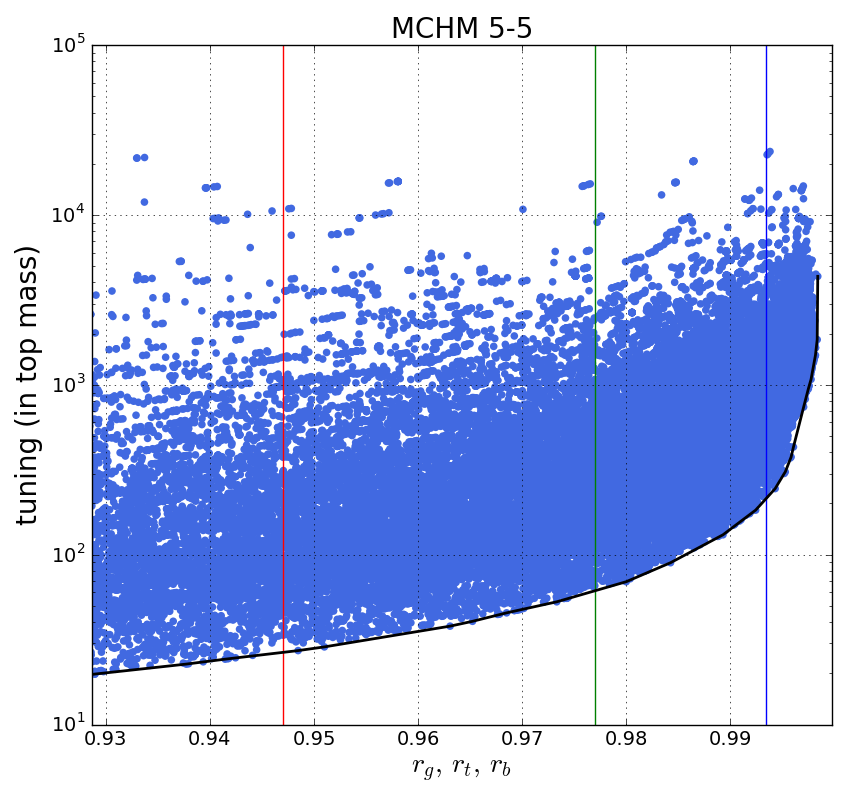}
\end{subfigure}%
\begin{subfigure}{0.5\textwidth}
\centering
\includegraphics[width=1\linewidth]{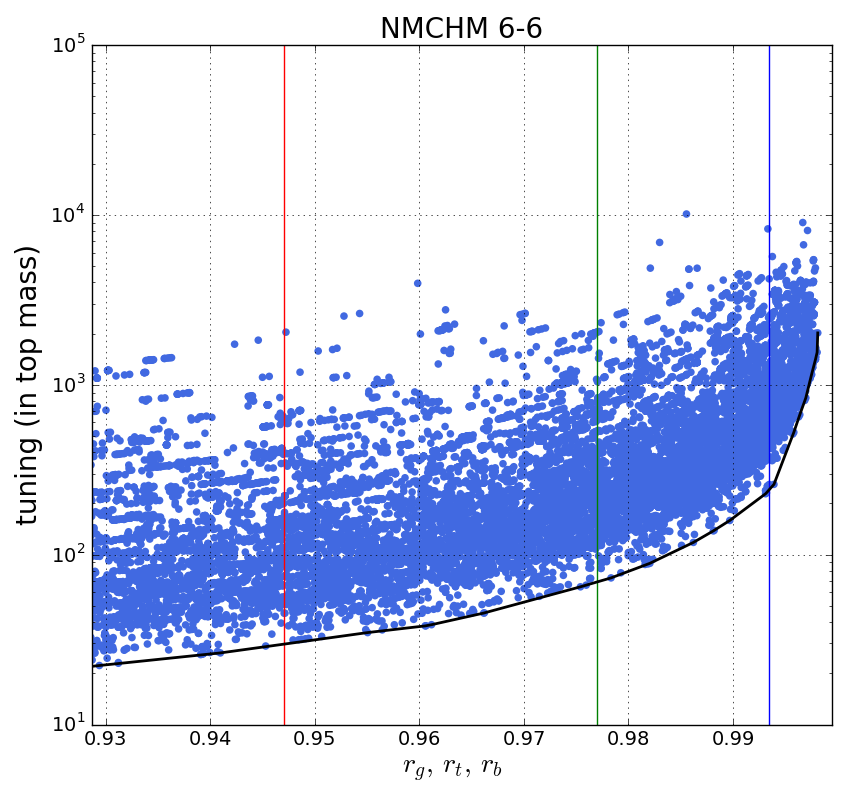}
\end{subfigure}\caption{\label{fig:first-tuning-top-vs-couplings} Comparison of the first-order tuning contribution from the top mass, in the Higgs-gluon, -top and -bottom coupling deviation. The red line shows the expected precision of a $250$ GeV ILC, green a $500$ GeV ILC, and blue a high-luminosity $1$ TeV ILC.}
\end{figure}

\begin{figure}[h]
\centering
\begin{subfigure}{0.5\textwidth}
\centering
\includegraphics[width=1\linewidth]{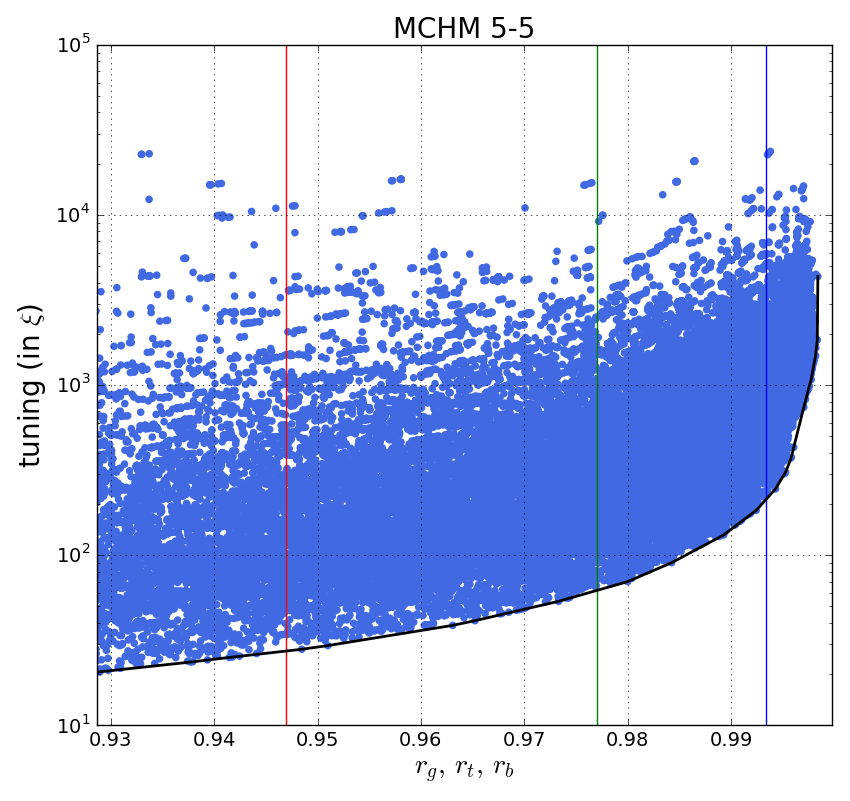}
\end{subfigure}%
\begin{subfigure}{0.5\textwidth}
\centering
\includegraphics[width=1\linewidth]{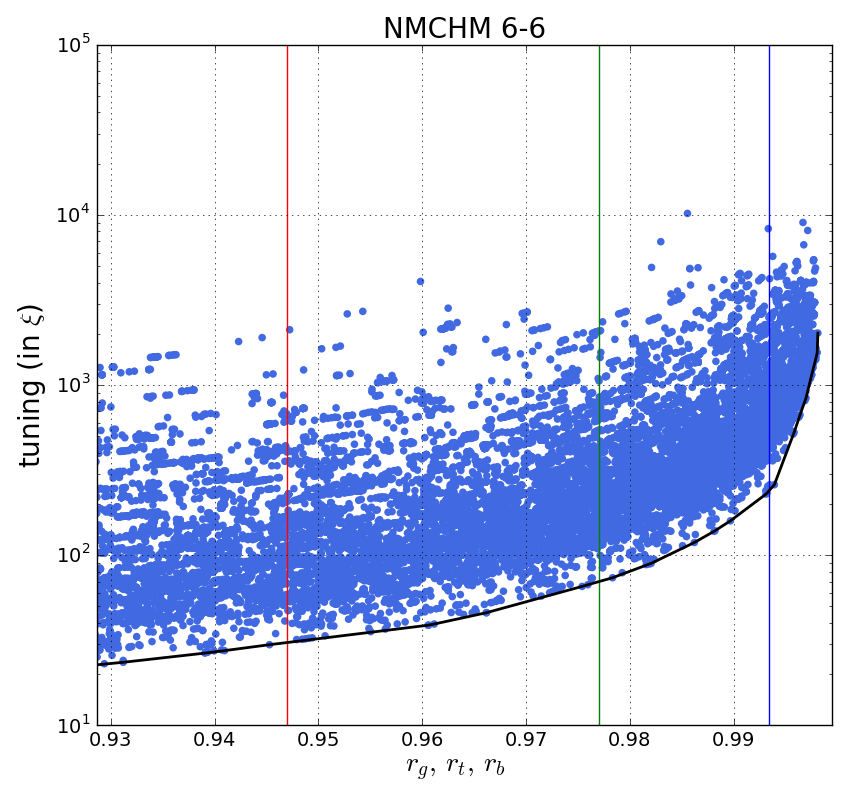}
\end{subfigure}\caption{\label{fig:first-tuning-vac-vs-couplings}Comparison of the first-order tuning contribution from the vacuum misalignment $\xi$, in the Higgs-gluon, -top and -bottom coupling deviation. The red line shows the expected precision of a $250$ GeV ILC, green a $500$ GeV ILC, and blue a high-luminosity $1$ TeV ILC.}
\end{figure}

To understand the different contributions to the higher-order tuning, we show in Figures~\ref{fig:first-tuning-higgs-vs-couplings} to~\ref{fig:first-tuning-vac-vs-couplings} various first-order tuning contributions (defined in Equation~\ref{fot}), again plotted as a function of the modification to the Higgs-gluon, Higgs-top and Higgs-bottom couplings. For any given value for the modification of the couplings, we see that tuning contribution from the Higgs mass is higher than that arising from the top mass and vacuum misalignment contributions. This can be understood from the leading-order relationship between each observable. By Equations \ref{eq:higgs_mass} and \ref{eq:top_mass}, for $\xi << 1$, $m_h \propto \xi^2$, while $m_t \propto \xi$ (recalling that $\frac{1}{f^2} = \frac{\xi}{v^2}$). Thus the first order tuning is expected to be $\nabla^{m_h} \sim 2\nabla^{m_t} \sim 2\nabla^{\xi}$, which agrees with Figures~\ref{fig:first-tuning-higgs-vs-couplings} to~\ref{fig:first-tuning-vac-vs-couplings}. 

\begin{figure}[h]
\centering
\begin{subfigure}{0.5\textwidth}
\centering
\includegraphics[width=1\linewidth]{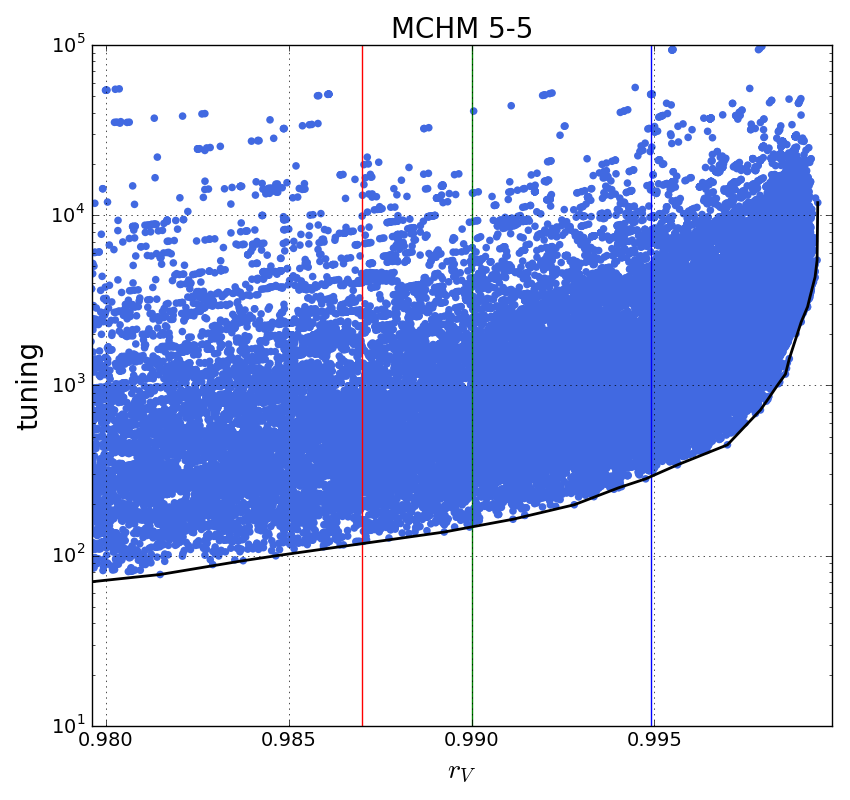}
\end{subfigure}%
\begin{subfigure}{0.5\textwidth}
\centering
\includegraphics[width=1\linewidth]{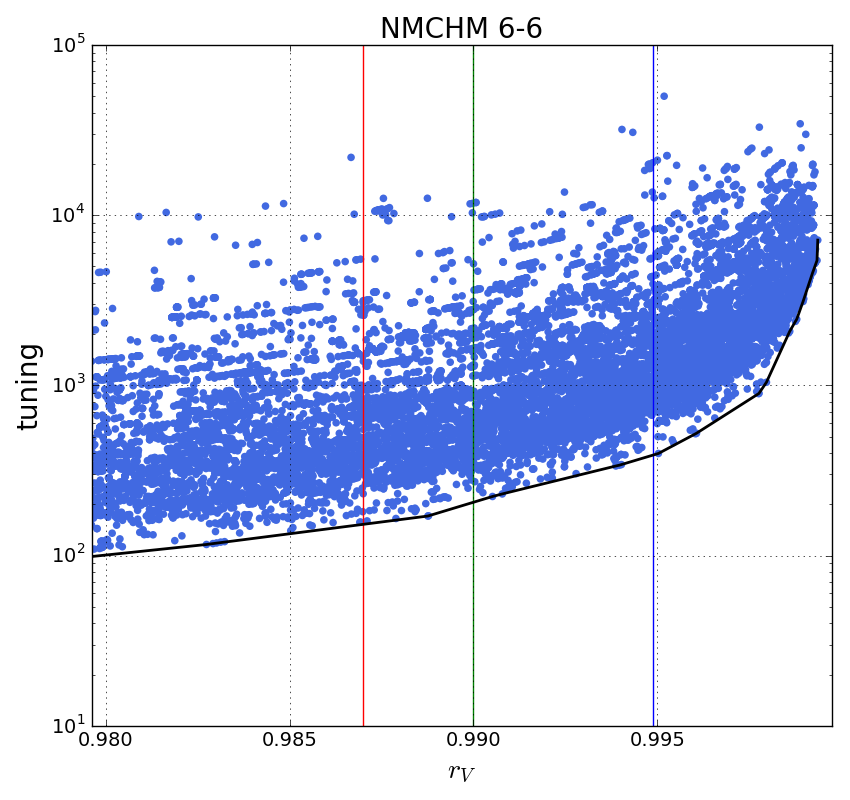}
\end{subfigure}\caption{\label{fig:tuning-higgs-vs-V} Comparison of higher-order tuning in the Higgs-vector boson coupling deviation, between the minimal and next-to-minimal models. The red line shows the expected precision of a $250$ GeV ILC, green a $500$ GeV ILC, and blue a high-luminosity $1$ TeV ILC.}
\end{figure}

\begin{figure}[h]
\centering
\begin{subfigure}{0.5\textwidth}
\centering
\includegraphics[width=1\linewidth]{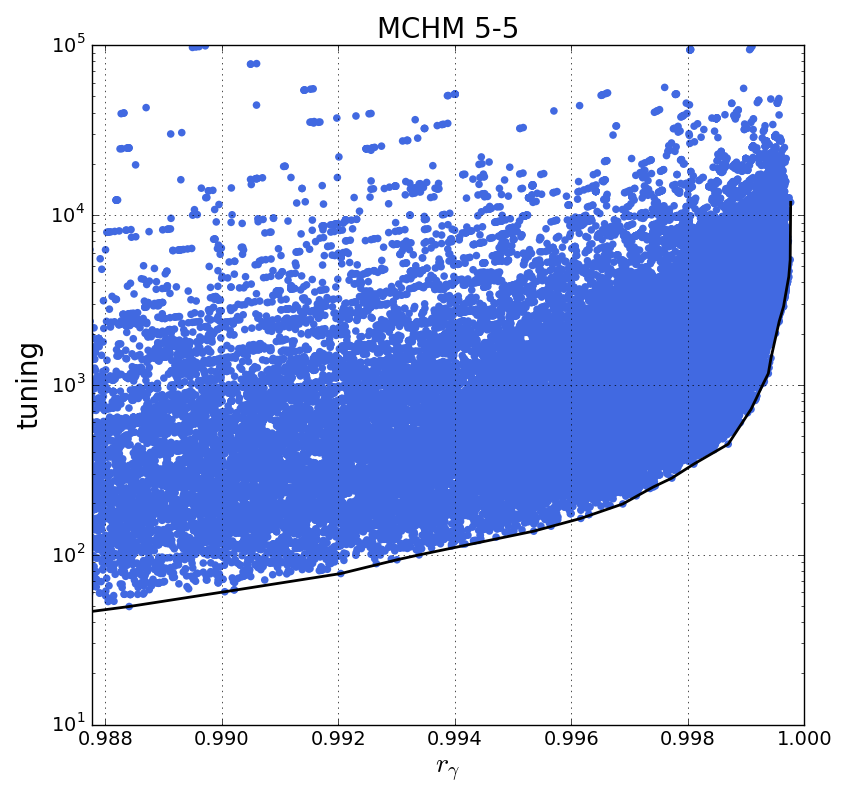}
\end{subfigure}%
\begin{subfigure}{0.5\textwidth}
\centering
\includegraphics[width=1\linewidth]{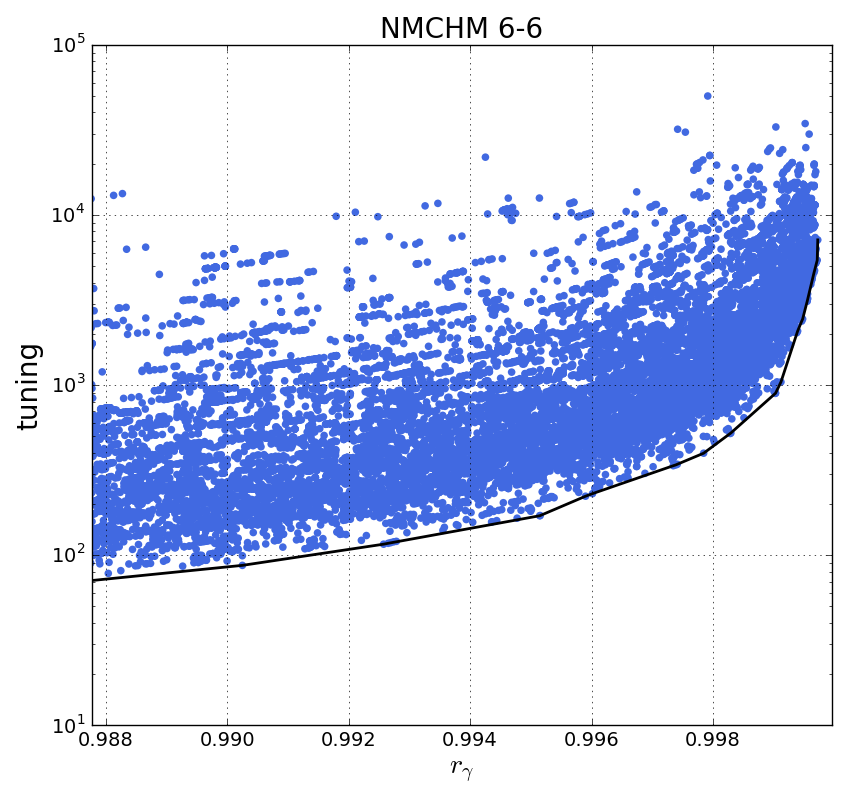}
\end{subfigure}\caption{\label{fig:tuning-higgs-vs-gamma} Comparison of higher-order tuning in the Higgs-photon (loop) coupling deviation, between the minimal and next-to-minimal models. Future ILC bounds are below the cut-off $f>800$ GeV.}
\end{figure}

In Figure~\ref{fig:tuning-higgs-vs-V}, we show the higher-order tuning as a function of the deviation of the Higgs-vector boson couplings. In this case, the impact of the future linear collider is not as pronounced, with a less pronounced increase in the fine-tuning even after the anticipated results of the high-energy, high-luminosity ILC. The situation is even worse for the Higgs-photon coupling (shown in Figure~\ref{fig:tuning-higgs-vs-gamma}), where the relative lack of precision of ILC measurements of $r_\gamma$ relative to the other couplings means that there is essentially no impact on the fine-tuning of the model expected from future measurements. This tells us that it is future measurements of the Higgs-gluon, Higgs-top and Higgs-bottom couplings that will be most important in disfavouring composite Higgs scenarios on aesthetic grounds.

\begin{figure}[h]
\centering
\begin{subfigure}{0.5\textwidth}
\centering
\includegraphics[width=1\linewidth]{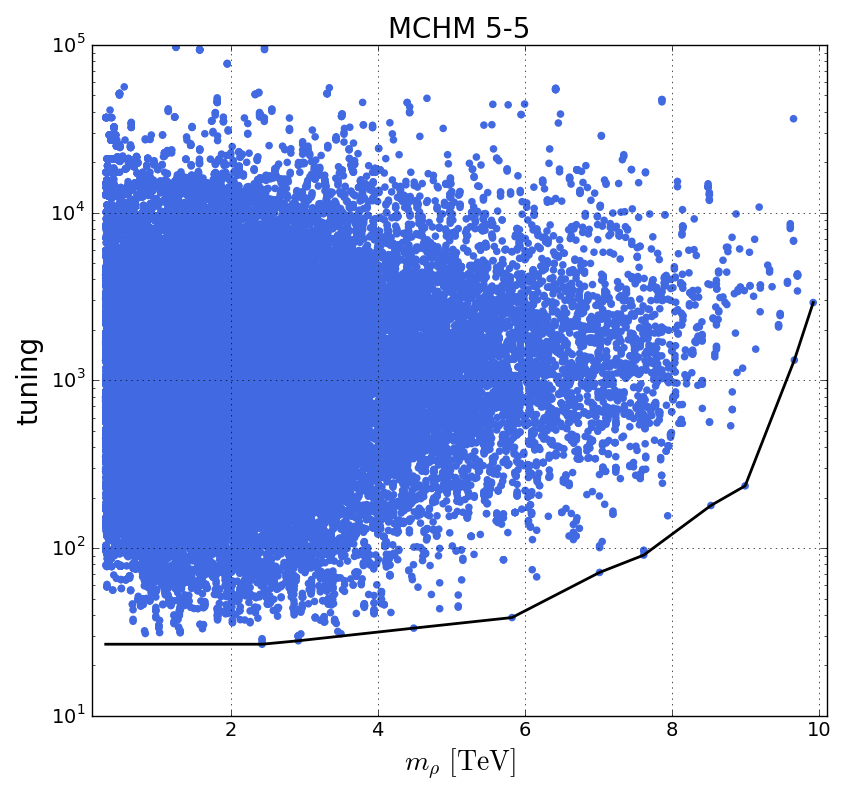}
\end{subfigure}%
\begin{subfigure}{0.5\textwidth}
\centering
\includegraphics[width=1\linewidth]{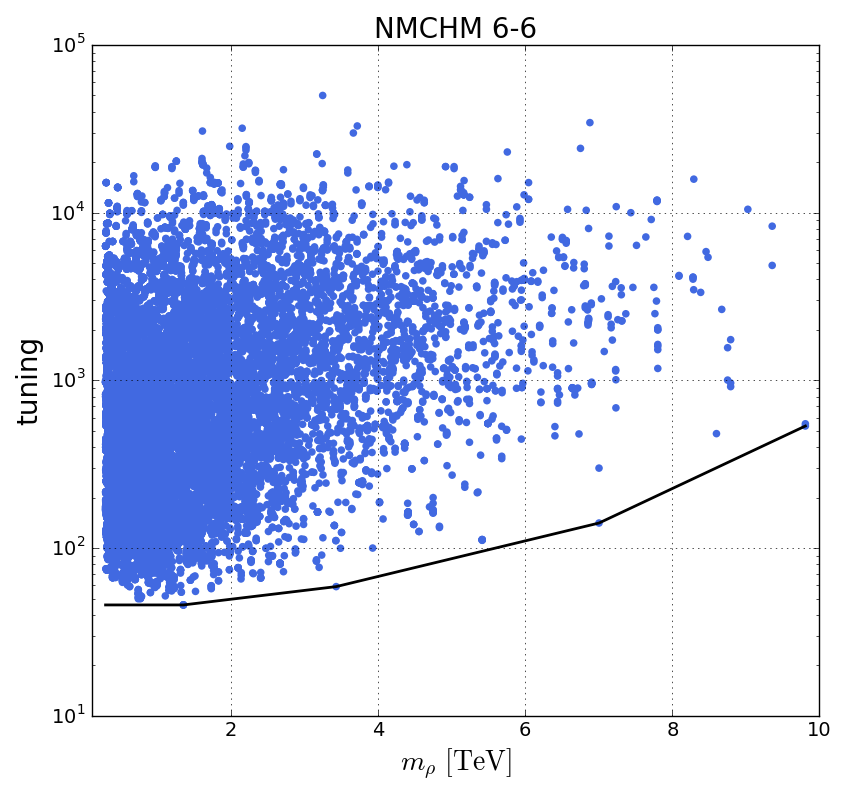}
\end{subfigure}\caption{\label{fig:tuning-vs-mrho} Comparison of the lightest vector resonance mass vs higher-order tuning, between models.}
\end{figure}

\begin{figure}[h]
\centering
\begin{subfigure}{0.5\textwidth}
\centering
\includegraphics[width=1\linewidth]{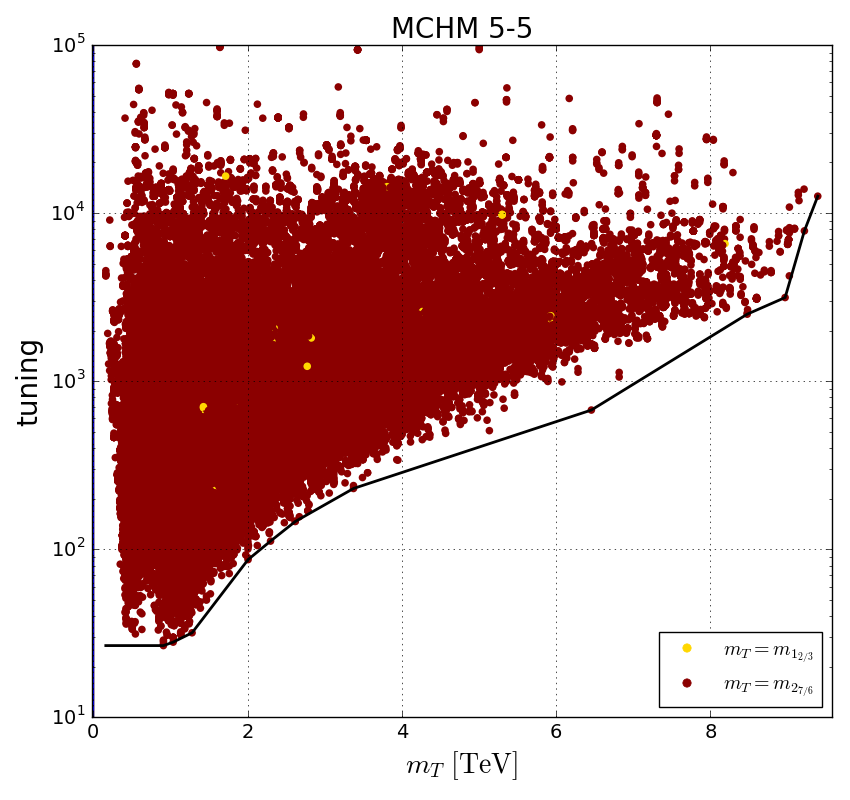}
\end{subfigure}%
\begin{subfigure}{0.5\textwidth}
\centering
\includegraphics[width=1\linewidth]{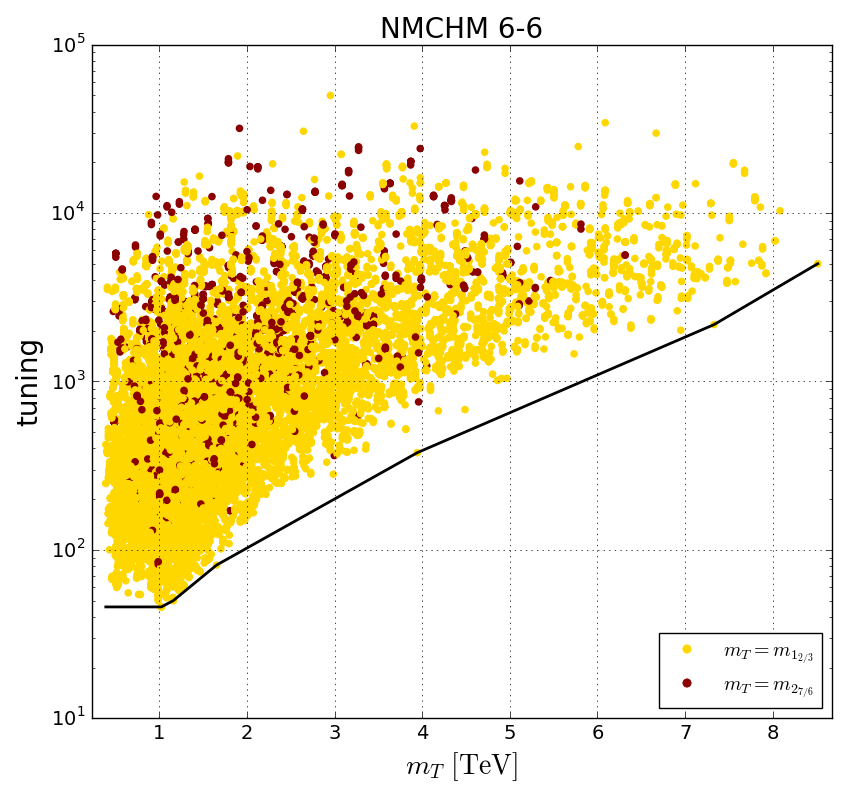}
\end{subfigure}\caption{\label{fig:tuning-vs-mT} Comparison of the lightest fermionic resonance mass vs higher-order tuning, between models. Note that the lightest resonance may be either the singlet (yellow) or doublet (maroon).}
\end{figure}

Measurements of the Higgs couplings are, of course, only one way to constrain the fine-tuning of the MCHM and NMCHM. One can also search directly for the fermion and vector resonances. In Figure~\ref{fig:tuning-vs-mrho}, we show the higher order tuning as a function of the lightest vector resonance mass, $m_\rho$. A lower bound on this mass would translate directly into a lower bound on the fine-tuning. In this case, the rise in fine-tuning with an increasing lower mass limit is less pronounced for the NMCHM, although one would have to have a fairly stringent lower bound to make this difference significant. A steeper rise is apparent in the plots of higher order tuning vs top partner mass $m_T$ shown in Figure~\ref{fig:tuning-vs-mT}, although there is not much difference in the behaviour in the NMCHM relative to the MCHM. Lower limits of around 5 TeV and 9.5 TeV can be expected after 3000 fb$^{-1}$ of 33 TeV and 100 TeV collisions at a future proton--proton collider, respectively~\cite{Gershtein:2013iqa,colliderReach,chala2018searches}, which will substantially increase the minimum fine-tuning of both the MCHM and NMCHM.

\begin{figure}[h]
\centering
\begin{subfigure}{0.5\textwidth}
\centering
\includegraphics[width=1\linewidth]{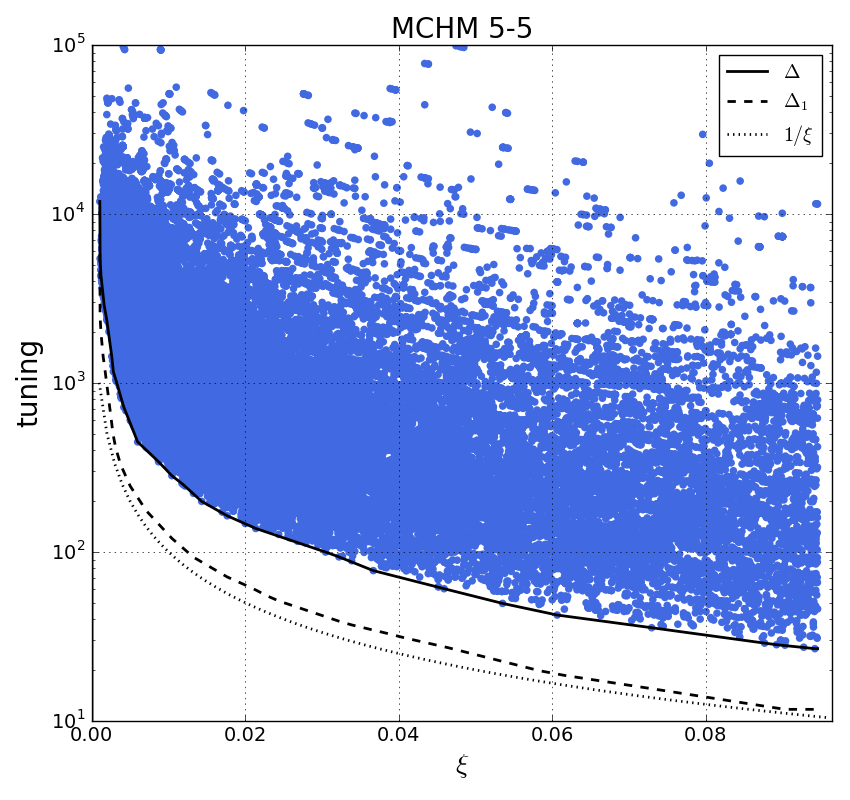}
\end{subfigure}%
\begin{subfigure}{0.5\textwidth}
\centering
\includegraphics[width=1\linewidth]{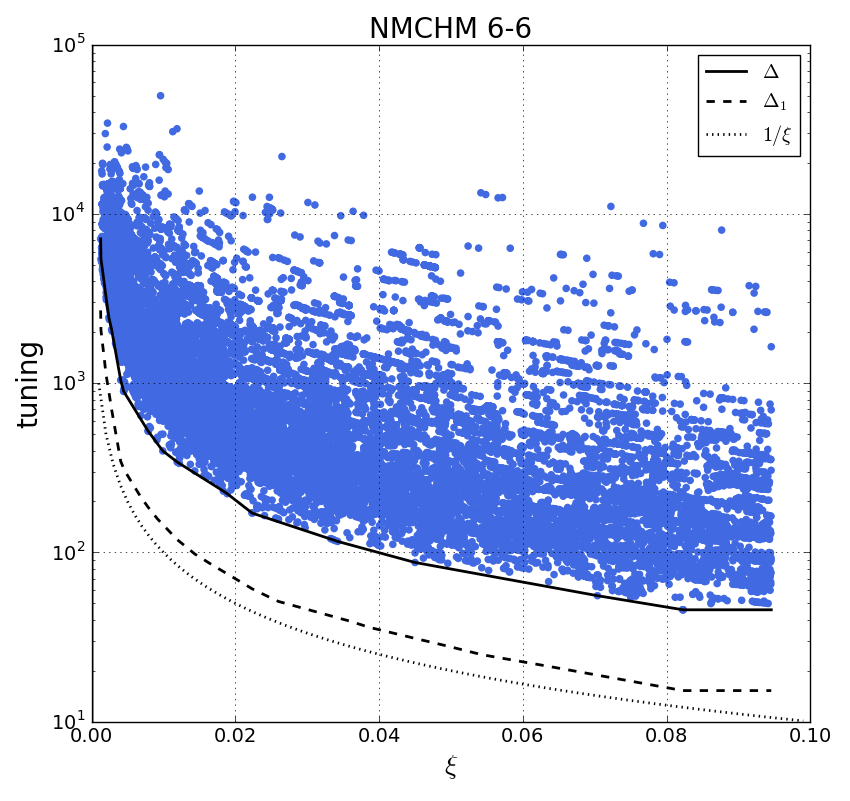}
\end{subfigure}\caption{Comparison of vacuum misalignment vs higher-order tuning, between models.}\label{fig:tuning_vs_xi}
\end{figure}

Finally, we show a comparison of our higher-order tuning measure with less sophisticated tuning measures in Figure~\ref{fig:tuning_vs_xi}, which shows the fine-tuning for the NMCHM as a function of the breaking scale ratio $\xi$. Our measure gives higher values for fine-tuning relative to the single tuning $\Delta_1$ as defined in Equation \ref{fot}, which is to be expected.

\begin{table} 
\begin{center}
\begin{tabular}{c | c | c | c}
Plot & $250\gev$ (red) & $500\gev$ (green) & $1\tev$ HL (blue) \\ \hline
$r_b$ & $5.3\%$ & $2.3\%$ & $0.66\%$\\
$r_Z$ & $1.3\%$ & $1.0\%$ & $0.51\%$\\
$r_\gamma$ & $18\%$ & $8.4\%$ & $2.4\%$
\end{tabular}
\end{center}
\caption{A selection of Higgs coupling deviation exclusion bounds, as predicted in \cite{tian2016measurement}. These are forecasts for ILC precision relative to the SM prediction.}\label{tab:ILC}
\end{table}
%

\clearpage

Before concluding,  let us briefly examine the behaviour of the singlet mass in our scan results. Apart from a dependence on the potential integral terms, it depends on the decay constant $f$ and the embedding angle of the right-handed top quark in the $SO(2)$ subgroup of $SO(6)$, $\theta$. We see that there is a critical point determined by the $\cos 2\theta$ factor in Equation \ref{eq:singlet_mass}, with the limits
\begin{align}
m_S^2 \rightarrow \begin{cases}
-\frac{c_1 - c_2 + c_3}{f^2}, & \text{ as } \sin\theta \rightarrow 1\\
0, & \text{ as } \sin\theta \rightarrow \frac{\sqrt{2}}{2} \approx 0.7
\end{cases}
\end{align}
The zero mass case corresponds to the right top embedding being $SO(2)$ symmetrical, leaving this group unbroken and the singlet as a true NGB. It has been shown in Reference~\cite{gripaios2009beyond} that two-loop contributions from the gauge sector will still give the singlet a small mass, appearing as an electroweak axion. This would be ruled out by experiment.

The $\sin\theta = 1$ limiting case is more interesting. Here, the elementary top quark does not couple with the singlet eigenstate, and Equations \ref{eq:gauge_lagrangian} and \ref{eq:fermion_lagrangian} become (considering only the subset of terms containing the $\psi$ field)
\begin{align}
\mathcal{L}_{\psi} \; \xrightarrow{\theta\rightarrow \pi/2} \; & (\partial \psi)^2 + \frac{g^2 f^2}{4} \sin^2 \frac{\varphi}{f} \cos^2 \frac{\psi}{f} W^2\\
&\stackrel{\psi \rightarrow -\psi}{=} \mathcal{L}_{\psi}
\end{align}

This $Z_2(\psi)$ symmetry is explored in Reference~\cite{Frigerio:2012uc}, where it is simply assumed. It requires all interactions to preserve $s$-number, which protects the scalar singlet from decay hence making it a suitable candidate for dark matter. In that work, the authors consider four regions of interest: low mass ($m_S < 50\gev$), resonant ($m_S \approx m_H/2$), cancellation ($m_S \sim \sqrt{\frac{\lambda}{2}} f$) and high mass ($m_S >> \sqrt{\frac{\lambda}{2}} f$). Here, $\lambda$ is the four-point coupling of $\psi,\varphi$, appearing in Equation \ref{potential}. In our notation, $\lambda \rightarrow c_1 - c_3$, since
\begin{align}
V(\psi,\varphi) \xrightarrow{s_\varphi^2,s_\psi^2 << 1} (c_1 + c_2 -c_3) \varphi^2 - (c_1 - c_3) \varphi^2 \psi^2 + c_3 \varphi^4 - c_3 \varphi^4 \psi^2
\end{align} 

\begin{figure}[h]
\centering
\begin{subfigure}{0.4\textwidth}
\centering
\includegraphics[width=1\linewidth]{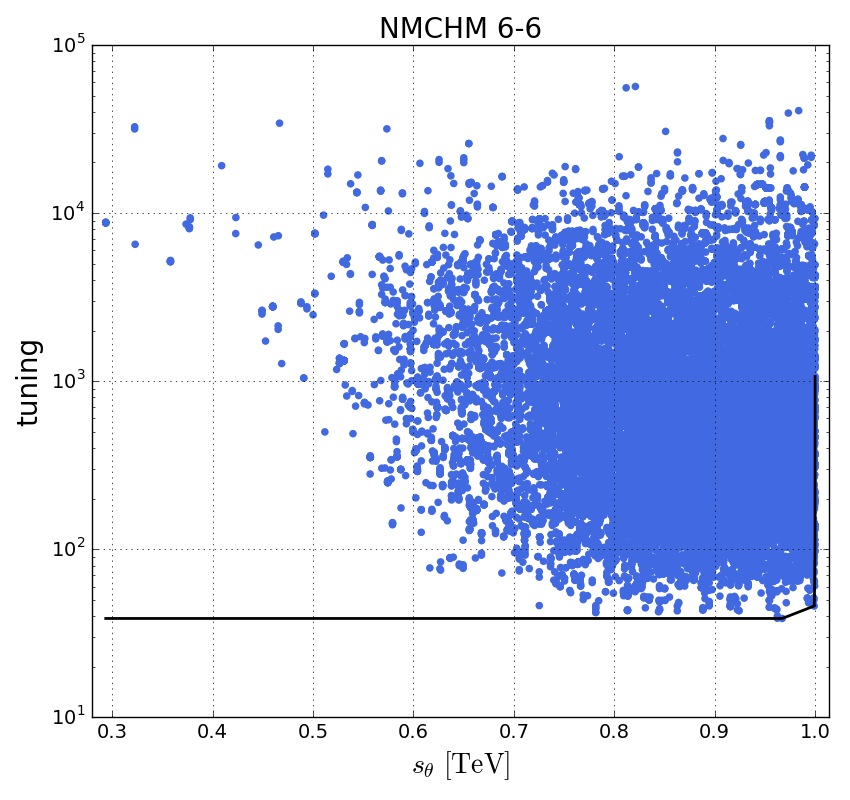}\caption{}
\end{subfigure}%
\begin{subfigure}{0.4\textwidth}
\centering
\includegraphics[width=1\linewidth]{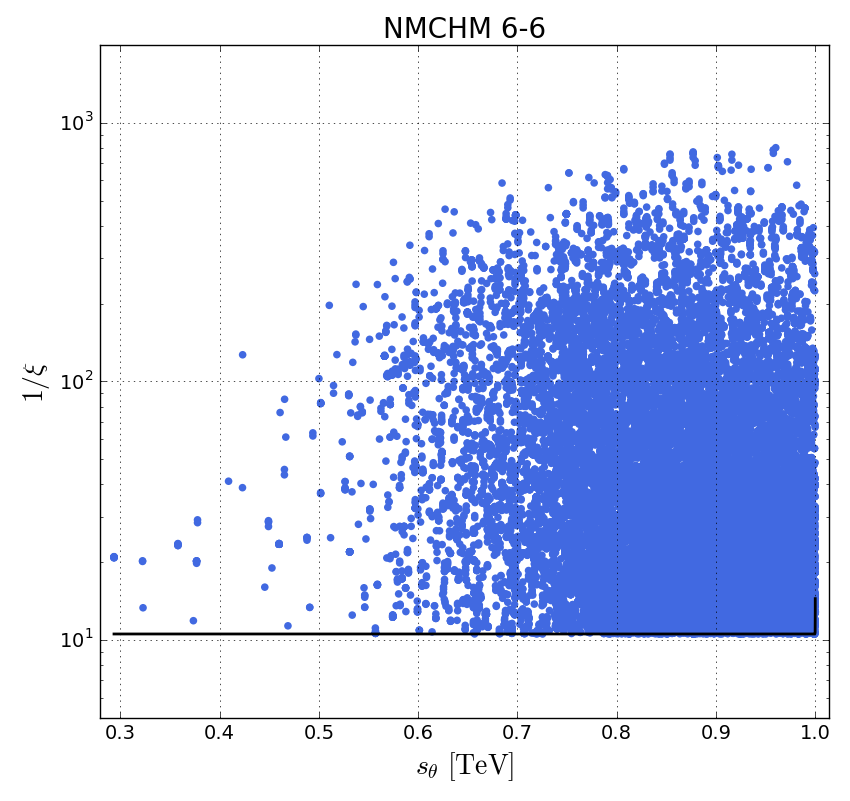}\caption{}
\end{subfigure}
\caption{The top quark mixing parameter $\sin\theta$ vs (top row left) higher order tuning and (top row right) naive tuning.}\label{fig:theta_vs_tuning}
\end{figure}

\begin{figure}[h]
\centering
\begin{subfigure}{0.6\textwidth}
\centering
\includegraphics[width=1\linewidth]{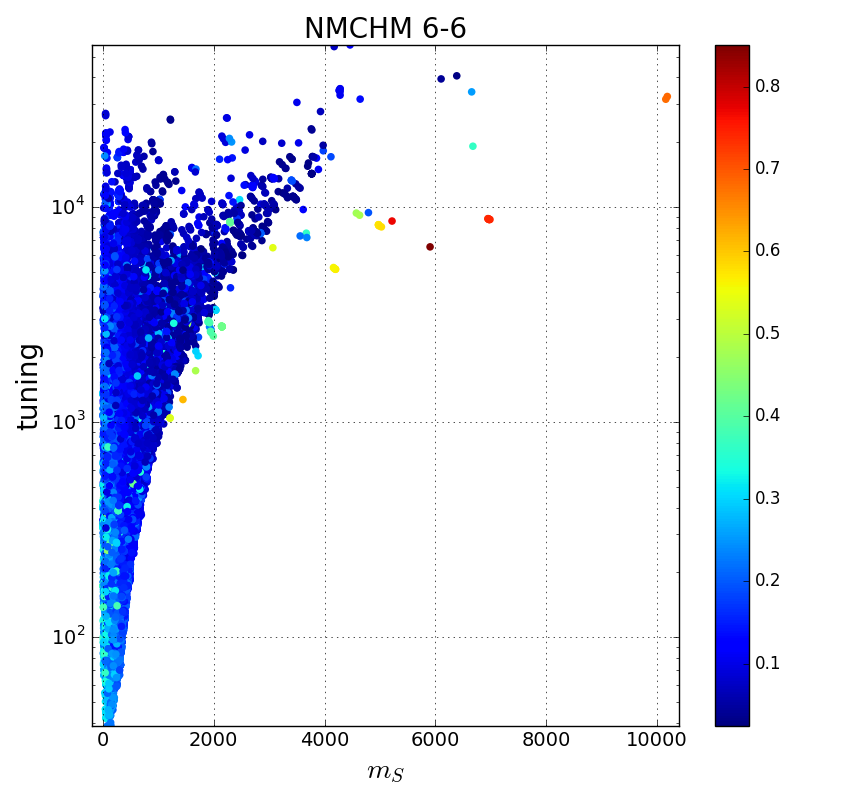}
\end{subfigure}
\caption{Mass of the singlet in GeV, with singlet-quark coupling deviation (as defined in Equation \ref{eq:coupling_expressions}) as the third dimension}\label{fig:mSvscoupling}
\end{figure}

In Figure~\ref{fig:theta_vs_tuning}, we show both our higher order fine-tuning measure, and the naive measure $1/\xi$, vs $\sin\theta$ for our selected scan points. We see that the NMCHM provides points with low fine tuning even as $\sin\theta\rightarrow 1$, and hence a dark matter candidate can easily emerge naturally within this framework. In Figure~\ref{fig:mSvscoupling}, we show our higher-order tuning measure vs the singlet mass, with the deviation of the singlet couplings to quarks and gluons from SM Higgs-like couplings shown on the $z$-axis (this deviation is defined in Equation~\ref{eq:coupling_expressions}). Higher values on the $z$-axis correspond to a stronger coupling between the singlet and quarks and gluons. We see that obtaining couplings as high as the SM Higgs requires a fine-tuning that is two orders of magnitude greater than the most natural coupling
scenario of small coupling.

It is instructive to separate our scan points into the region that has $\theta < \pi/4$,  and that which has $\theta > \pi/4$. Moving from one choice to the other requires a change in the sign of the $c_1 - (c_2 + c_3)s_\theta^2$ term to guarantee a real singlet mass. Specifically, by removing $f$ as a factor in the mass term using the solution for $\xi$, we get the regions in terms of only the integral expressions
\begin{align}
\text{Region 1: } && \theta \in \{0,\pi/4\}, && \implies && \frac{c_2^2 s_\theta^4 - (c_1 - c_3 s_\theta)^2}{2 c_3} > 0\nonumber \\
\text{Region 2: } && \theta \in \{\pi/4, \pi/2\}, && \implies && \frac{c_2^2 s_\theta^4 - (c_1 - c_3 s_\theta)^2}{2 c_3} < 0\label{eq:regions}
\end{align} 

\begin{figure}[h]
\centering
\begin{subfigure}{0.5\textwidth}
\centering
\includegraphics[width=1\linewidth]{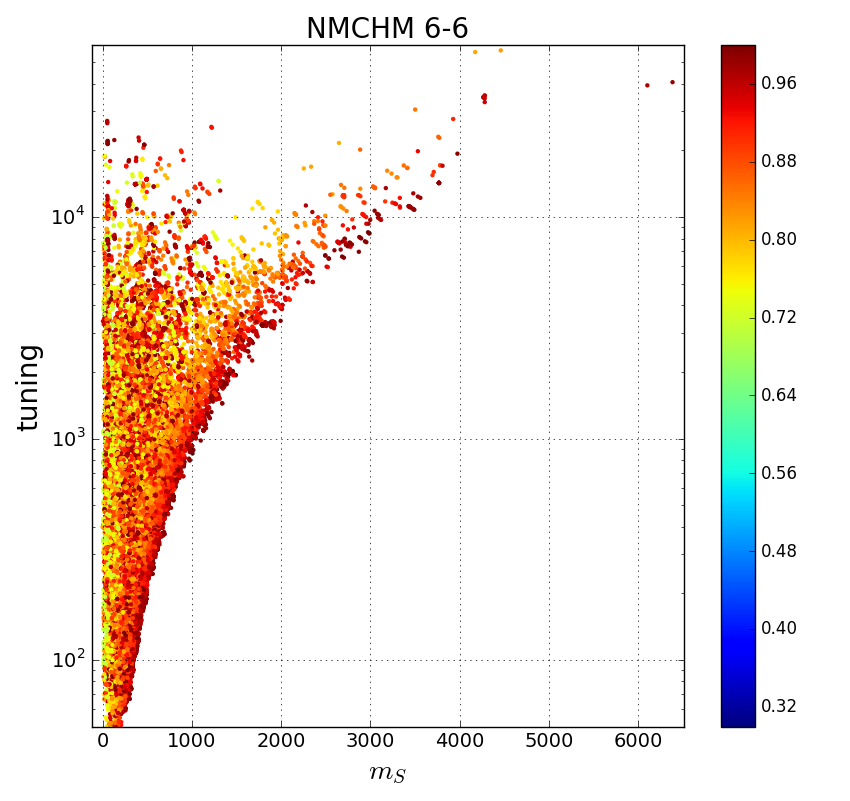}\caption{}\label{fig:mS_vs_tuning1}
\end{subfigure}%
\begin{subfigure}{0.5\textwidth}
\centering
\includegraphics[width=1\linewidth]{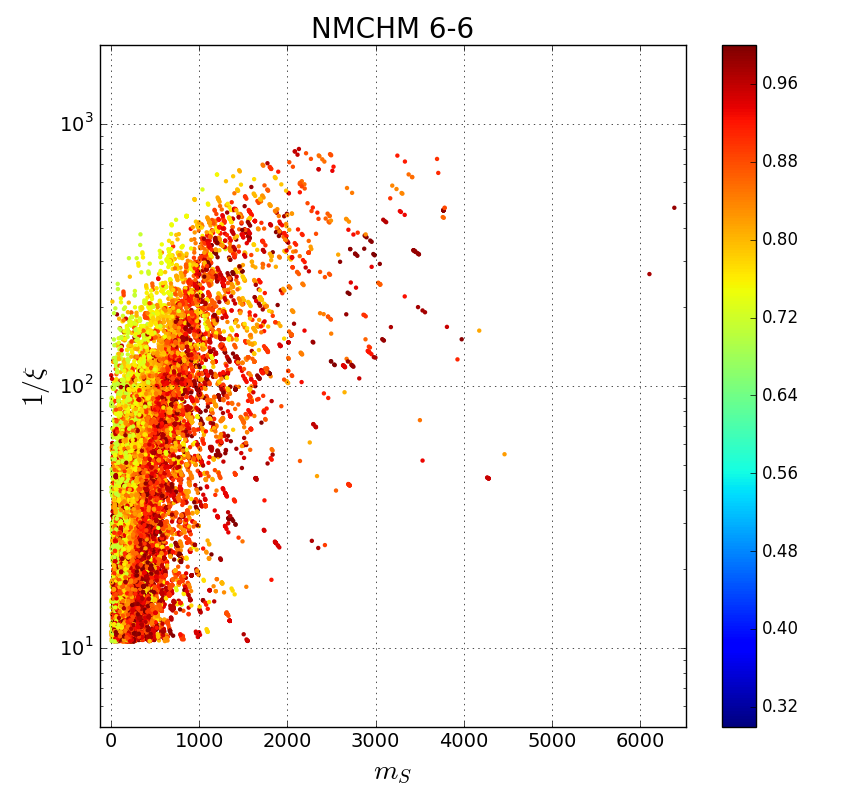}\caption{}\label{fig:mS_vs_xi1}
\end{subfigure}
\caption{The singlet mass (in GeV) vs (a) higher order tuning and (b) naive tuning with $\sin\theta$ as the third dimension, for points with $\theta \in \{\pi/4, \pi/2\}$.\label{fig:mS-R2}}
\end{figure}

\begin{figure}[h]
\centering
\begin{subfigure}{0.5\textwidth}
\centering
\includegraphics[width=1\linewidth]{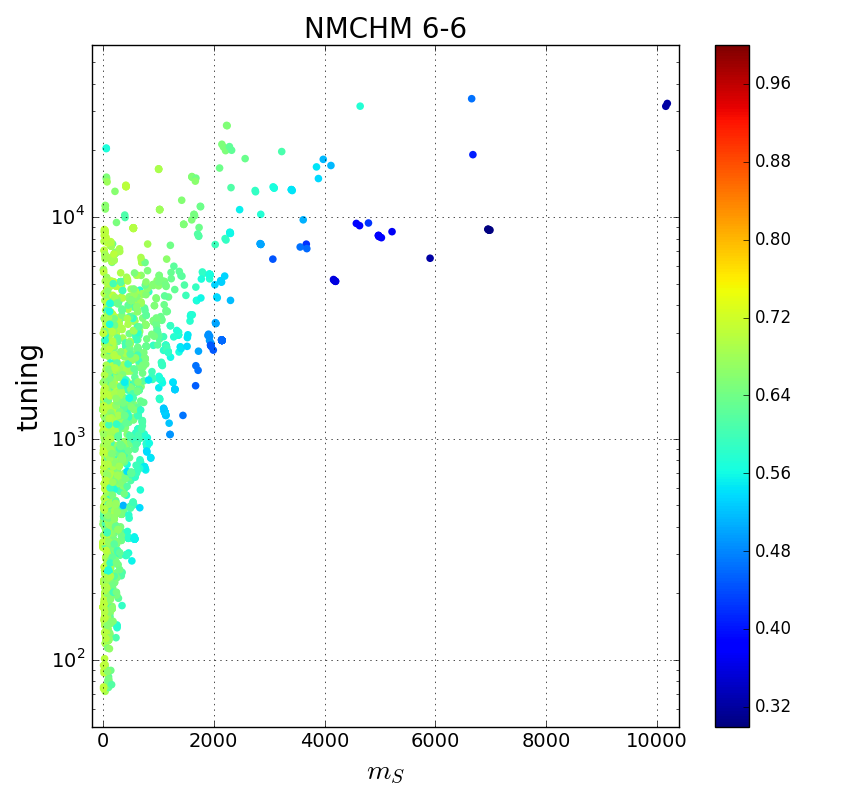}\caption{}\label{fig:mS_vs_tuning1}
\end{subfigure}%
\begin{subfigure}{0.5\textwidth}
\centering
\includegraphics[width=1\linewidth]{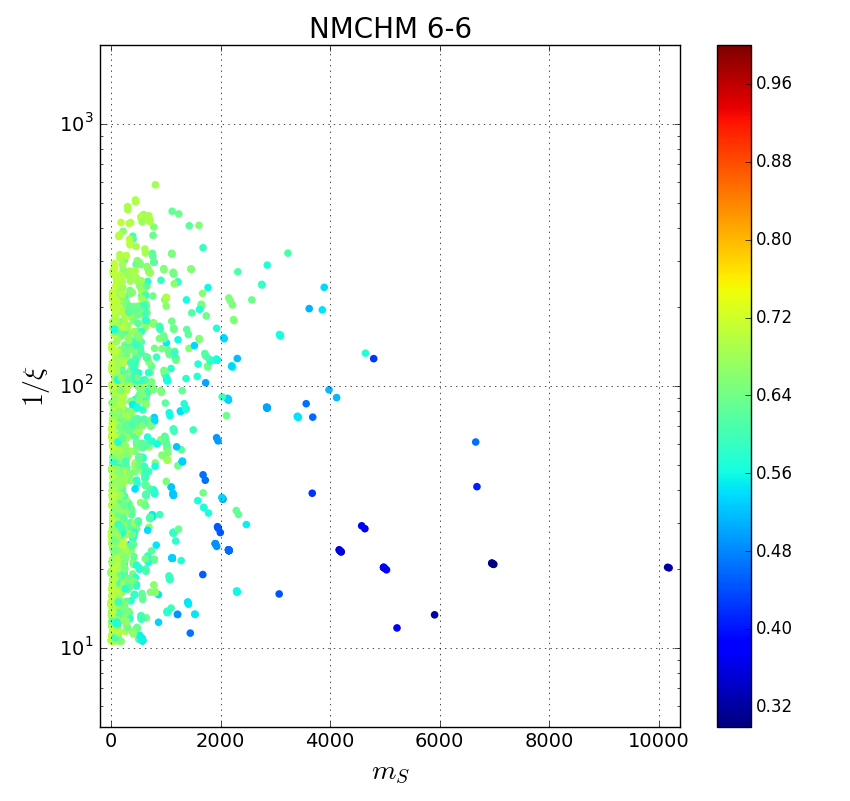}\caption{}\label{fig:mS_vs_xi1}
\end{subfigure}
\caption{The singlet mass (in GeV) vs (a) higher order tuning and (b) naive tuning with $\sin\theta$ as the third dimension, for points with $ \theta \in \{0,\pi/4\}$.\label{fig:mS-R1}}
\end{figure}

In Figure~\ref{fig:mS-R2}, we show our higher-order tuning measure, and the naive tuning measure, vs $m_S$ for points with $\theta \in \{\pi/4, \pi/2\}$, indicating that the points of lowest tuning have sin$\theta$ values close to 1. This implies that the $Z_2$ symmetry exists to stabilise a dark matter candidate. Equivalent plots for our $ \theta \in \{0,\pi/4\}$ points are shown in Figure~\ref{fig:mS-R1}, in which the contour of lowest fine tuning now exists such as to minimise sin$\theta$. In both cases, the features are pronounced only when considering the higher-order tuning measure which counts multiple contributions to the total fine-tuning properly. We note that the lowest fine-tuning overall is usually encountered for points with $ \theta \in \{0,\pi/4\}$, but that there is not a large difference between the overall fine-tuning vs mass for the two $\theta$ regions. A final note regarding the dark matter candidacy of the singlet; the natural limit of $\sin\theta \approx 1$ suggested by Figure \ref{fig:mS-R2} is based only on considerations of SM mass values. It is not an indication of fine tuning based on cosmological values. Indeed, to achieve the correct relic density of the DM candidate, one may need to be arbitrarily close to the $Z_2$ limit. In this sense, enforcing the limit could be considered a separate source of fine tuning. It is beyond the scope of this paper to provide relic density limits on the singlet-fermion coupling terms. Suffice it to say that given the effective next-to-minimal model, particularly for higher singlet masses, the $\sin\theta \approx 1$ region is preferred by particle mass tuning considerations, and one would be well-motivated to search for UV completions that included this $Z_2$ symmetry explicitly.

\begin{figure}[h]
\centering
\begin{subfigure}{0.4\textwidth}
\centering
\includegraphics[width=1\linewidth]{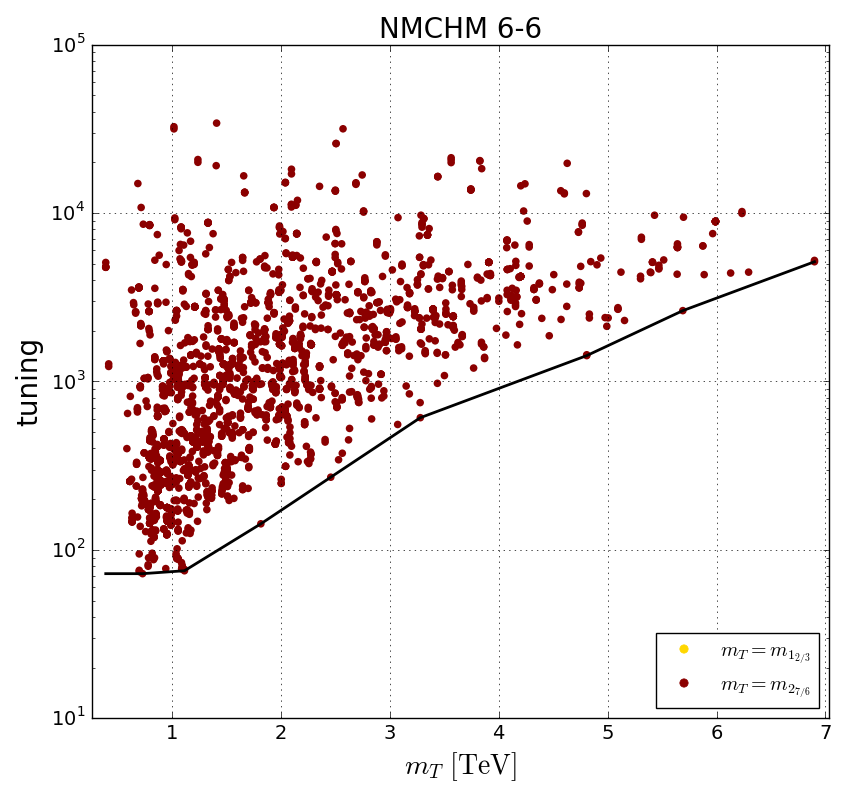}
\end{subfigure}%
\begin{subfigure}{0.4\textwidth}
\centering
\includegraphics[width=1\linewidth]{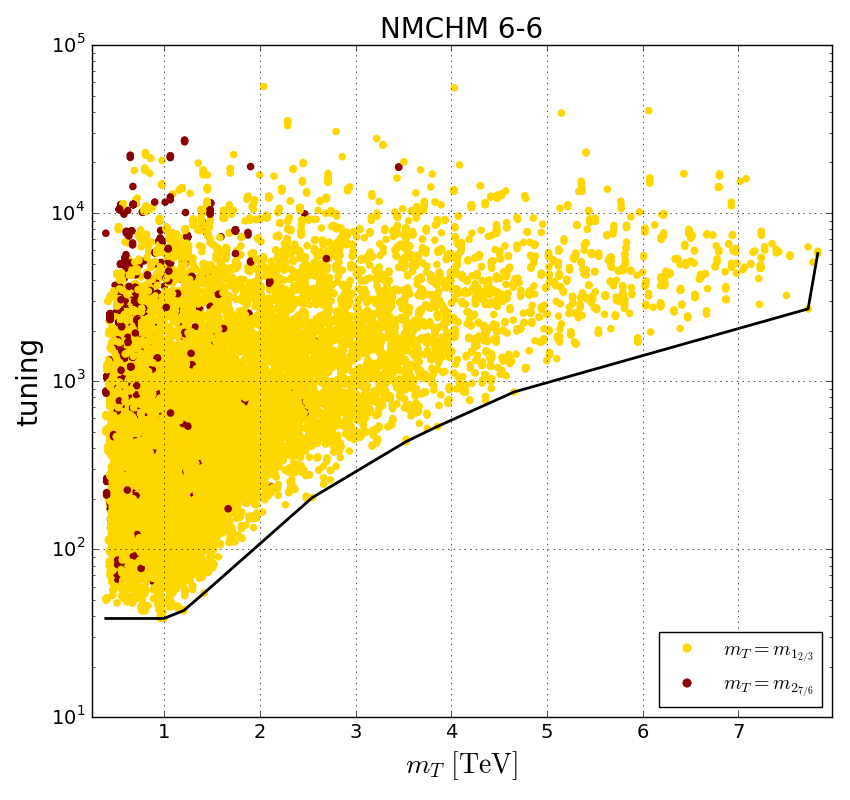}
\end{subfigure}
\caption{Top partner masses vs. full tuning, broken into region 1 (left) and region 2 (right), as defined by Equation \ref{eq:regions}.}\label{fig:Tvsth}
\end{figure}

In Figure \ref{fig:Tvsth}, we show the higher-order tuning vs the lightest top partner mass, showing by the colour of each point which of the two top partners is the lightest. The left-hand plot contains only the points with $ \theta \in \{0,\pi/4\}$, whilst the right-hand plot shows the points with $\theta \in \{\pi/4, \pi/2\}$. Our results suggest that a collider observation of a lightest top partner with hypercharge $2/3$ will always allow the identification $\theta \in \{\pi/4, \pi/2\}$ under the assumption that the NMCHM is a valid explanation, whereas any observation of the hypercharge will allow the identification of the $\theta$ region for a lightest top partner mass in excess of $3.5\tev$. In turn, this would allow one to infer the singlet's phenomenology, if one were to construct the model with the minimum fine-tuning.

\section{Conclusions}
\label{sec:conclusions}
We have performed a detailed comparison of the fine-tuning of the NMCHM and the MCHM, with, in each case, partially composite third generation quarks embedded in the fundamental representation of the relevant global symmetry group. Using a new scanning technique, differential evolution, we were able to accurately map the regions of the whole parameter space that simultaneously give the correct SM Higgs mass, Higgs VEV and SM top quark mass, whilst minimising our novel measure of fine-tuning that correctly counts multiple sources of fine-tuning. By showing the fine-tuning as a function of the resonance masses and deviations to the Higgs couplings, we were able to assess the impact that future collider measurements on these quantities will have on the minimum fine-tuning available in either model.

In general, we find little difference in the behaviour of the MCHM and NMCHM, beyond a slight increase in the fine-tuning of the NMCHM which results from our measure penalising the extra complexity of the latter model. As a benchmark, the MCHM had a minimum tuning of $\Delta \sim 26$, while the NMCHM had a minimum tuning of $\Delta \sim 45$. Future high-luminosity measurements of the Higgs coupling to third-generation quarks and gluons at a 1 TeV ILC can be expected to increase the fine-tuning of the MCHM and NMCHM by approximately one order of magnitude relative to the best present-day precision of only $\sim 9\%$ (for gluons \cite{ATLAS-CONF-2018-031,PhysRevD.98.030001}), as could a bound on the lightest top partner mass of $\approx 4-5$~TeV. In the NMCHM, we find that the ability of the extra scalar to act as a dark matter candidate, through the realisation of a $Z_2$ symmetry that prevents it from decaying, does not come with a fine-tuning penalty. On the contrary, the $Z_2$ symmetric limit of the theory is associated with parameter values that are amongst the least finely-tuned. 

\section*{Acknowledgements}

The work of MJW is supported by the Australian Research Council Future Fellowship FT140100244. DTM and AGW. are supported by the ARC Centre of Excellence for Particle Physics at the Terascale (CoEPP) (CE110001104) and the Centre for the Subatomic Structure of Matter (CSSM). DTM is supported by an Australian Government Research Training Program (RTP) Scholarship.


\appendix

\section{Fermion Representation Expressions}
\label{expressions}

Here we present the explicit low-energy expressions derived from the high-energy Lagrangian (Equation \ref{eq:fundamental_lagrangian}). All broken form factors can be expressed in the formulas

\begin{align}
\hat{\Pi}[m_1,m_2,m_3] &= \frac{(m_2^2 + m_3^2 - p^2)\Delta^2}{p^4 - p^2(m_1^2 + m_2^2 + m_3^2)+m_1^2 m_2^2},\\
\hat{M}[m_1,m_2,m_3] &= \frac{m_1 m_2 m_3 \Delta^2}{p^4 - p^2(m_1^2 + m_2^2 + m_3^2) + m_1^2 m_2^2}
\end{align}
In terms of these formulas, the broken factors are given by
\begin{align}
\hat{\Pi}_0^{q_L} &= \hat{\Pi}[m_{T},m_{\tilde{T}},m_{Y_T}], & \hat{\Pi}_1^{q_L} &= \hat{\Pi}[m_{T},m_{\tilde{T}},m_{Y_T} + Y_T] - \hat{\Pi}[m_{T},m_{\tilde{T}},m_{Y_T}], \\
\hat{\Pi}_0^{u_R} &= \hat{\Pi}[m_{\tilde{T}},m_T,m_{Y_T}], & \hat{\Pi}_1^{u_R} &= \hat{\Pi}[m_{\tilde{T}},m_{T},m_{Y_T} + Y_T] - \hat{\Pi}[m_{\tilde{T}},m_{T},m_{Y_T}], \\
\hat{M}_0^{u} &= \hat{M}[m_{T},m_{\tilde{T}},m_{Y_T}], & \hat{M}_1^{u} &= \hat{\Pi}[m_{T},m_{\tilde{T}},m_{Y_T} + Y_T] - \hat{\Pi}[m_{T},m_{\tilde{T}},m_{Y_T}].
\end{align}
These broken form factors contribute to the full form factors present in the electroweak EFT Lagrangian \ref{eq:fermion_lagrangian}
\begin{align}
\Pi_0^q &= \frac{1}{y_{t_L}^2} + \hat{\Pi}_0^{q_L}, &\Pi_1^{q_1} &= \hat{\Pi}_1^{q_L}, \\
\Pi_0^u &= \frac{1}{y_{t_R}^2} + \hat{\Pi}_0^{u_R} + s_{\theta}^2 \hat{\Pi}_1^{u_R}, & \Pi_1^u &= -2\hat{\Pi}_1^{u_R},\\
M_1^u &= \hat{M}_1^u.
\end{align}
The Higgs-singlet potential is presented to quartic order in section \ref{sec:GB_Vacuum}. We repeat it here
\begin{align}
V(\varphi,\psi) \approx c_1 s^2_\varphi c^2_\psi + c_2 s^2_\varphi (s^2_\theta - c^2_\theta s^2_\psi) - c_3 s^2_\varphi c^2_\psi (c^2_\theta s^2_\varphi s^2_\psi + s^2_\theta c^2_\varphi)
\end{align}
with the integral terms given by
\begin{align}
c_1 &= -N_c \int\frac{d^4 p}{(2\pi)^4}\frac{\Pi_1^{q_1}}{\Pi_0^q} + V(h)_\textnormal{gauge}, & c_2 &= -N_c \int\frac{d^4 p}{(2\pi)^4} \frac{\Pi_1^u}{\Pi_0^u}, \\
c_3 &= -N_c \int\frac{d^4 p}{(2\pi)^4}\frac{(M_1^u)^2}{\left(\Pi_0^q + s^2_\varphi c^2_\psi \Pi_1^{q_1}/2\right)\left(\Pi_0^u + s^2_\varphi c^2_\psi s^2_\theta \Pi_1^u/2\right)},
\end{align}
where 
\begin{align}
V(h)_\textnormal{gauge} & \approx \frac{9}{64\pi^2}\frac{g_0^2}{g_\rho^2}\frac{m_\rho^4 (m_{a_1}^2 -m_\rho^2)}{m_{a_1}^2 - m_\rho^2 (1 + g_0^2/g_\rho^2)}\ln\left[ \frac{m_{a_1}^2}{m_\rho^2(1+ g_0^2/g_\rho^2)}\right]
\end{align}
The masses of the fermion partners are given by the poles and roots of the following form factors
\begin{align}
\Pi_0^q (m_{\bm{2}_{1/6}}^2) = 0 && \frac{1}{\Pi_0^{q_L} (m_{\bm{2}_{7/6}}^2)} = 0 && \Pi_0^u (m_{\bm{1}_{2/3}}^2) = 0 
\end{align}
The following are the leading order deviations from the SM Yukawa couplings, as defined by Equation \ref{coupling} 
\begin{align}
r_{htt,hbb,hgg} = \frac{1-2\xi}{\sqrt{1-\xi}} && r_{stt,sbb,sgg} = \sqrt{\frac{\xi}{1-\xi}}\cot\theta \nonumber\\
r_{hVV} = \sqrt{1-\xi} && r_{h\gamma\gamma} = \frac{A_1 r_{hVV} + \frac{4}{3}A_{1/2} r_{htt}}{A_1 + \frac{4}{2}A_{1/2}} \label{eq:coupling_expressions}
\end{align}
where $A_1 \approx -8.324$ and $A_{1/2} \approx 1.375$.

\bibliographystyle{unsrt}
\bibliography{tex/refs}

\end{document}